\documentclass[aps,pre,preprint,groupedaddress,
,superscriptaddress,showpacs,a4paper,floatfix]{revtex4-1}

\usepackage{amsmath,amssymb,graphicx,graphics,bm}
\usepackage{hyperref}
\usepackage{graphics,graphicx}
\usepackage{amsmath,amssymb}

\begin{document}

\title{Fluid-particle interactions and fluctuation-dissipation relations I -
General linear  theory and basic fluctuational patterns}

\author{Massimiliano Giona}
\email{massimiliano.giona@uniroma1.it}
\affiliation{Facolt\`{a} di Ingegneria, La Sapienza Universit\`{a} di Roma,
via Eudossiana 18, 00184, Roma, Italy}

\author{Giuseppe Procopio}
\email{giuseppe.procopio@uniroma1.it}
\affiliation{Facolt\`{a} di Ingegneria, La Sapienza Universit\`{a} di Roma,
via Eudossiana 18, 00184, Roma, Italy}
 \author{Chiara Pezzotti}
\email{chiara.pezzotti@uniroma1.it}
\affiliation{Facolt\`{a} di Ingegneria, La Sapienza Universit\`{a} di Roma,
via Eudossiana 18, 00184, Roma, Italy}

\date{\today}

\begin{abstract}
The article provides a unitary and complete solution to the fluctuation-dissipation relations
for particle hydromechanics in a generic
fluid, accounting for the hydrodynamic fluid-particle interactions (including
arbitrary memory kernels in the description of dissipative
and fluid inertial  effects) in linear  hydrodynamic regimes, via the concepts of
fluctuational patterns.
This is achieved
by expressing the memory kernels as a linear superposition of 
exponentially decaying modes. 
Given the structure of the  interaction with the internal
degrees of freedom, and assuming  the representation of the thermal force as a superposition
of modal contributions, the fluctuation-dissipation relation follows simply
from the moment analysis of the corresponding Fokker-Planck equation,
imposing the condition that at equilibrium all the internal
degrees of freedom are uncorrelated with particle velocity.
Moreover, the functional structure of the  resulting
equation of motion 
corresponds to the principle
of complete decoupling amongst  the internal degrees of freedom.
The theory is extended to the case of confined geometries, by generalizing previous results
including the effect of fluid inertia.
\end{abstract}

\maketitle

\section{Introduction}
\label{sec1}

Fluctuation-dissipation theory represents a major field in statistical physics
connecting the deterministic (particle or field) interactions of 
a physical system (e.g. a micrometric particle in
a fluid) with the characterization 
 of stochastic thermal fluctuations  that should  be taken into account in 
order to  describe  its statistical properties at thermal equilibrium \cite{gen1,gen2}.
Pioneered  by the research of Einstein, Langevin, Smoluchoswki  \cite{einstein,langevin,smolu} 
on the physics of Brownian
motion -  expressed in its simplest form, i.e. for a spherical particle
subjected to the Stokesian drag and to a random thermal force - 
the concept of fluctuation-dissipation relations evolved in a manifolds of different
branches and meanings:  the extension to electric fluctuations \cite{el1,el2,el3},
the connection to macroscopic properties of materials (e.g. the susceptivities) \cite{susc1,susc2},
the application to  quantum fluctuations \cite{quantum1,quantum2},
the  statistical physics of fields (e.g. statistical hydrodynamics)
 \cite{landau,hydrofield1,hydrofield2} and generalized field equations \cite{ht1}, the extension
to non-equilibrium conditions (mostly considering   steady states) 
\cite{ht2,mm1,mm2}, the analysis of molecular simulation experiments in a statistical
physical perspective \cite{mm3,mm4}.

In this work, we focus on the classical problem of the thermal motion of a 
micrometric particle in a generic fluid of arbitrary complex linear rheological
behaviour, that corresponds to the generalization
of the seminal work by Kubo \cite{kubo,kubolibro}, accounting also for the
detailed representation of the fluid-particle interactions \cite{kim, brenner, makosko}.
This problem is  not only crucial in the theory of Brownian motion, as recent
accurate experiments at short time  scales have revealed \cite{expbm1,expbm2,expbm3,expbm4,expbm5},
but it plays also a fundamental role in the design of microfluidic systems 
\cite{procopiofluids,vendittifluids},  of chemical separation processes
at microscale and of analytical chemical microassays \cite{analchem1,analchem2,analchem3}. It is therefore of central importance both in theory and in practice.

The Kubo theory of fluctuation-dissipation originated as the
 generalization of the
Einstein-Langevin picture, where the Stokesian drag was extended to encompass dissipative memory
effects \cite{kubo,kubolibro}. This has represented a major 
advance in statistical
physics, as  it has permitted to include in a simple  description the class of models
introduced by Zwanzig \cite{zwanzig,zwanziglibro}, and referred to as Generalized
Langevin Equations (GLE).
Regarded in the light of a hydrodynamic theory of fluid-particle interactions, the Kubo
extension solve just one half of the problem and probably the  more peculiar one.
The experiments on Brownian motion even for Newtonian fluids (such as water or acetone at room
temperature)  \cite{expbm1,expbm2,expbm3}
have clearly shown that fluid inertial effects (referred to,
 in fluid dynamics, as 
the Basset force and the added mass contribution \cite{landau}) are the main responsible
for the long-term tails  in the particle velocity autocorrelation function.
The inclusion of these effects in a generalized fluctuation-dissipation theory
for  particle motion in a fluid is the basic prerequisite in order to 
predict the outcome of highly resolved Brownian experiments both in space and time,
and to   improve the description of fluid-particle
interactions in  Stokesian dynamics and suspension theory  \cite{brady,suspensions}
beyond the
the range of applicability of the Stokes regime.

In  linear hydrodynamic regimes (Stokes flows, time-dependent Stokes flows \cite{kim,brenner}), the motion of particle of mass $m$ is  described by the
evolution equation for the particle velocity $v(t)$ of the form
\begin{equation}
m \, \frac{d v(t)}{d t}= F_{f \rightarrow p}[v(t)] + R(t)
\label{eq1_1}
\end{equation}
where ${F}_{f \rightarrow p}[v(t)]={\mathcal L}[v(t)]$  represents
 the mean-field hydrodynamic
force exerted by the fluid onto the particle 
and is a linear functional of $v(t)$ (and of its time derivative),
and $R(t)$  the random force. Throughout this work we adopt the notationally simpler
scalar representation for particle dynamics (corresponding to
the description of a spherical Brownian particle in the free space). 
The vectorial extension and the
inclusion of the tensorial character of the hydromechanic kernels are  addressed in 
Section \ref{sec9} (in connection with problems in confined
geometries) and in Appendix \ref{app1} (as a generalization of the theory).
Moreover, we consider exclusively the translatory motion of a rigid particle,
leaving rotational and roto-translatory effects to a specific and further analysis.

Following Kubo \cite{kubo,kubolibro}, the
 fundamental problems in the fluctuation-dissipation theory
of particle motion in a fluid are essentially two:
\begin{itemize} 
\item the determination of the velocity autocorrelation
function $C_{vv}(t)= \langle v(t)  \, v(0) \rangle_{\rm eq}$;
\item  the determination
of the correlation properties of thermal force $R(t)$.
\end{itemize}
where $\langle \cdot \rangle_{\rm eq}$ indicates the expected
value with respect to the equilibrium probability measure.

The first problem can  be solved in general, invoking the principle
of uncorrelation of the thermal force $R(t)$   from $v(0)$ 
for $t\geq 0$,
\begin{equation}
\langle R(t) \,  v(0) \rangle_{\rm eq} =0
\label{eq1_2}
\end{equation}
that extends the original intuition by Paul Langevin \cite{langevin}.
Eq. (\ref{eq1_2}) is referred to by some authors as the {\em causality
principle} \cite{bedeaux1,mazur}.
Essentially, it means that the future evolution of $R(t)$, i.e., for $t \geq 0$
is completely independent of the  state of the velocity variable
at any previous time ($t=0$).
This corresponds to the situation in which $R(t)$ is not influenced
by the previous history of the  particle velocity, and consequently
$v(t)$ behaves passively with respect to $R(t)$ without altering
its future evolution.
For this reason, eq. (\ref{eq1_2}) can be properly indicated as
{\em the principle of fluctuational independence for passive
particles}, or simply as the Langevin  condition. Conversely, if eq. (\ref{eq1_2}) is not fulfilled,
particle dynamics may have an active influence on the outcome
of the fluctuational force. This issue is addressed in \cite{part3}.

Enforcing eq. (\ref{eq1_2}), the equation for the velocity
autocorrelation function $C_{vv}(t)$ follows straightforwardly
\begin{equation}
m \, \frac{d C_{vv}(t)}{d t}= {\mathcal L}[C_{vv}(t)]
\label{eq1_3}
\end{equation}
equipped with the initial condition $C_v(0)=\langle v^2 \rangle_{\rm eq}$,
that  represents the basic result of the Linear Response Theory 
(LRT for short) \cite{kubolibro}.
Conversely,
the determination of the thermal force  $R(t)$ and its statistical properties
has been fully addressed
solely in the  some particular cases
as discussed below.

In  the terminology introduced by Kubo \cite{kubolibro} eq. (\ref{eq1_3}) corresponds
to the {\em Fluctuation-Dissipation relation of the first kind} (FD1k, for short),
while the problem of deriving the  correlation
properties  of the thermal force $R(t)$ is named by
Kubo as the {\em  Fluctuation-Dissipation relation of the second kind} (FD2k, for short).
Moreover, Kubo observed that FD2k is not a simple problem as the separation
of the force into a hydrodynamic, ${\mathcal L}[v(t)]$ in eq. (\ref{eq1_2}),
and a random force $R(t)$ is a complex problem in statistical physics (see \cite{kubolibro} p. 37).
This is the reason why other alternatives to FD2k are possible, namely the
description of particle thermal motion  via
 a comprehensive stochastic hydrodynamic velocity field, encompassing both the effects (hydrodynamic and fluctuational) in a single
entity \cite{gionaklages}.

In the light of the accurate description of particle motion at microscales,
be it associated with Brownian motion theory or with  particle dynamics
in a microfluidic device, it  is convenient to introduce
a further relation, referred to as the {\em Fluctuation-Dissipation
relation  of the third kind} (FD3k), corresponding to the explicit
representation of the thermal force, consistent with FD1k and FD2k,
in terms of elementary stochastic processes, which represents the
fundamental theoretical tool to infer prediction about Brownian
and microparticle dynamic properties beyond the
analysis of  its correlation properties of the second-order, addressing
also microlocal  and regularity properties  of trajectories
and velocity realizations (see \cite{part3}).

Many developments have been carried out to extend the 
classical framework pioneered by Kubo, in order  to account for a wide range of dynamic phenomena, such as those observed in non-Markovian, semi-Markovian or 
anomalous systems.

The fundamental role of the dissipative memory kernel has been pointed out in the analysis of many biological environments, in which the form of kernel itself is a key feature in the description of the physics. For instance, the integration of semi-Markov models has been effective to describe molecular motors in viscoelastic environments, such as kinesin motion \cite{Knoops2018,Lisy2016}, while coarse-grained modeling of non-Markovian dynamics is particularly effective in capturing the dynamics of colloidal suspensions and active matter \cite{Jung2018,Li2023}. 

Advances in theoretical formulations of GLE have been achieved through the exploration of hydrodynamic memory effects and non-Gaussian diffusive behaviors. For instance, the memory kernel formulation has been revisited to account for geometrical inhomogeneities and non-classical shapes of the particles \cite{VillegasDiaz2021}. In parallel, the analysis of non-Gaussian diffusion in heterogeneous media has bring the necessity for a unifying framework to combine distinct mechanisms of heterogeneity \cite{Lanoiselee2019}.

From a methodological perspective, novel solutions to the GLE have been proposed to compute thermodynamic quantities such as entropy production and work in systems with memory \cite{Terlizzi2020}. These works are complemented by explicit extensions of fluctuation-dissipation theorems for systems experiencing hydrodynamic backflow and confined geometries, which incorporate memory kernels with non-trivial decay behavior \cite{Lisy2016,Tothova2016,Maes2013}.

Since experimental results on Brownian motion further underscored the necessity of dissipative and viscoelastic effects into predictive models \cite{Ginot2022}, different studies have investigated both their impact on the fluctuation-dissipation relation \cite{Fox1977,Doerries2021,Yuyu2015}, and their interactions with external perturbations \cite{Allegrini2007,Holtz2018} or anomalous leading effects \cite{Baczewski2013,Baldovin2019,Goy2020}.

All these references show how in the purely dissipative case the analysis developed by Kubo is completely self-consistent \cite{kubo,kubolibro} and, out of it, it is possible
to provide an explicit representation of $R(t)$ \cite{goychuk} (see also \cite{mod1,mod2,mod3,mod4}).

Moreover,  the  
general case of a complex viscoelastic fluid  for which
both dissipative and fluid inertial contributions are simultaneously accounted
for has not been considered. 
More precisely, Bedeaux and Mazur \cite{bedeaux_inertial}
proposed a formal solution to this problem in a Newtonian fluid, in terms of the
Fourier transform of the autocorrelation function of $R(t)$, but this solution
does not correspond to any  feasible stochastic process. This has been discussed in
\cite{gionaklages}, and we arrive to the same conclusion in section \ref{sec7} from
a totally different perspective.

The scope of this work, that represents the first part of  a more comprehensive
analysis and extension of fluctuation-dissipation  relations 
(see \cite{part2,part3}),
is to provide a general and unitary solution of FD3k for particle motion
in a fluid in the  presence
of arbitrary linear hydrodynamic regimes and fluid-particle interactions,
both in the free space and in confined systems.
This unitary picture is achieved via the concept of {\em fluctuational patterns}
and their additive algebra that follows directly from the Langevin
condition. This  permits  further  and natural generalizations
of fluctuation-dissipation  theory, removing the Langevin condition eq. 
(\ref{eq1_2}), and addressing the regularity properties of velocity dynamics
in a comprehensive perspective, see \cite{part3}.

In developing the theory we make use of  two main tools: (i) the
representation of hydromechanic memory kernels  in terms of
exponentially decaying modes, and (ii) moment analysis
of the associated extended particle dynamics.

The representation of the
memory kernel in terms of
exponentially decaying modes
is  the standard rheological description of complex fluids in the linear
hydrodynamic regime \cite{makosko,rheol}, usually
referred to  as the Prony series
expansion \cite{prony1,prony2}. 
In the statistical mechanical literature it represents  the key assumption
for deriving a Markovian embedding out of a memory dynamics.

The extension of this approach to
the fluid inertial contribution has been  developed in \cite{gionavisco}.
Moreover, following  \cite{gionavisco}, we enforce the physical property
that the fluid inertial kernel is a bounded function of time. 
This property stems from the
finite propagation velocity of the shear stresses. In this framework, the singularity of the
Basset kernel at  time $t=0$ should be regarded as the rather unphysical byproduct 
of the purely Newtonian
constitutive equations (instantaneous Newtonian approximation) \cite{part3}, as  
the boundedness of this kernel is a mathematical property of
any viscoelastic fluids \cite{gionavisco}. In point of fact,
 any fluid, that 
for practical hydrodynamic purposes is regarded as  Newtonian, 
possesses a small but non-vanishing relaxation
time (which is order of 1 ps for water at room temperature \cite{water}).
Recently, Giona et al. \cite{gpp_localrepr}
argued that this approximation might represent the
consequence of  the local stationary and linear nature
of fluid-particle interactions (principle of local realizability)
associated with the  modes of the  hydrodynamic field interacting with the
particle.

Moment analysis is a classical tool in the study of stochastic 
processes \cite{gardiner}. The application of moment analysis
to particle hydromechanics in generic fluid and flow conditions
is particularly cogent as it permits to unveil many subtle
and fundamental issues of fluctuation-dissipation theory: (i) the
complete independence of the internal degrees of freedom at equilibrium,
associated with the detailed Langevin conditions stemming from eq. (\ref{eq1_2}),
leading to the concept of fluctuational patterns (Section \ref{sec8}),
(ii) the extension of the theory to the confined case  with its peculiarities
as regards the formulation of a thermodynamically consistent mean-field
hydromechanic force. For this reason, moment analysis is
introduced starting from the
simplest problem, i.e. the purely dissipative case of Section \ref{sec4}
(for which the representation of $R(t)$ is well known \cite{goychuk}),  
as a fil-rouge connecting all  the cases of increasing
hydrodynamic complexity in a  single methodological  perspective.

The article is organized as follows. Section \ref{sec2} introduces the 
hydromechanic problem, and Section \ref{sec3}
addresses the modal representation of the hydrodynamic kernels. 
Section \ref{sec4} develops
the FD3k representation in the simplest case of purely dissipative memory without
any fluid inertial effect. 
Albeit the results obtained coincide with those derived in \cite{goychuk},
the analysis of this simple case permits us to introduce  and 
describe the  approach to FD3k 
that is  extended to the  general  case in the subsequent Sections. 
Section \ref{sec5} briefly reviews a simple but important representation issue 
associated
with fluid inertial effects, and Section \ref{sec6} 
develops the analysis of the FD3k relation
in the most general hydrodynamic conditions  
where both dissipative and fluid inertial
memory  effects are considered. 
Section \ref{sec7} develops an explicit representation
of the stochastic force $R(t)$ related to fluid inertial
effects deriving FD2k in this case. 
From   the expression of the correlation function of the inertial stochastic
perturbation, the regularity  issue of the inertial kernel at  time
$t=0$ becomes
analytically evident.
Section \ref{sec8}  reframes the results obtained in the previous Sections 
in a unitary
respective via the concepts of fluctuational patterns.
Up to \ref{sec8} the analysis is focused on a particle in a free space,
neglecting the constraints deriving from
the presence of solid walls, such as in microfluidic channels
\cite{procopiofluids,brenner}, and their hydromechanic implications
(position dependent memory kernels).
This extension is provided in Section \ref{sec9} that generalizes the
results recently obtained in \cite{gppfd} to include fluid inertial
effects. The analysis of this
problem puts fluctuation-dissipation theory in a completely
new light,  from the sheer description of thermal fluctuations
to  the  necessary theoretical
tool to infer   a thermodynamically consistent 
 mean-field  hydromechanics via the analysis of its response in a thermal
fluctuational field at equilibrium.
The
extension of the theory to vectorial models and to the case of nonisotropic 
hydrodynamic interactions, and some numerical examples are
left to the Appendices.

\section{Particle hydromechanics}
\label{sec2}
The motion of a spherical Brownian particle  of mass $m$ 
at constant temperature $T$ in a quiescent fluid,
considered by Einstein and Langevin,  can
be described by the simple stochastic equation for the particle velocity $v(t)$
\begin{equation}
m \, d v(t) =  - \eta \, v(t) \, d t +  \sqrt{2 \, k_B \, T \, \eta} \, d w(t)
\label{eq2_1}
\end{equation}
where $\eta$ is the friction factor, $k_B$ the Boltzmann constant and 
$dw(t)$ the increment of a Wiener process in the time interval $(t,t+dt)$.
Introducing the dimensionless quantities
\begin{equation}
\widetilde{v}= \frac{v}{\sqrt{\langle v^2 \rangle_{\rm eq}}} = \sqrt{\frac{m}{k_B \, T}} \, v
\, , \quad \widetilde{t}= \frac{t}{t_{\rm diss}}= \frac{\eta}{m} \, t
\label{eq2_2}
\end{equation}
where $t_{\rm diss}=m/\eta$ is the characteristic
dissipation time,
eq. (\ref{eq2_1}) attains the nondimensional form
\begin{equation}
d \widetilde{v}= - \widetilde{v} \, d \widetilde{t} + \sqrt{2} \, d w(\widetilde{t})
\label{eq2_3}
\end{equation}
where $d w(\widetilde{t})$, as above, corresponds  to the increment
of a Wiener process in the interval $(\widetilde{t},\widetilde{t}+d \widetilde{t})$.
Consequently, $\langle \widetilde{v}^2 \rangle_{\rm eq}=1$.
Henceforth, we will use this nondimensional formulation unless otherwise stated (as
in  Sections \ref{sec8} and \ref{sec9}), by replacing, for notational simplicity, $\widetilde{v}$ and $\widetilde{t}$ with $v$ and $t$,  thus
$\langle v^2 \rangle_{\rm eq}=1$.

The Einstein-Langevin equation (\ref{eq2_1}) stems from the assumption
that the fluid model  is Newtonian and it  can be described by means of the instantaneous Stokes
equation. The timeless and instantaneous 
response of the fluid  corresponds 
to the occurrence of a $\delta$-correlated white-noise thermal force $R(t)= \sqrt{2 \, k_B \, T \, \eta} \, \xi(t)$ in eq. (\ref{eq2_1}),
where $\xi(t)=d w(t)/dt$ is the distributional
derivative of a Wiener process, $\langle \xi(t) \, \xi(t^\prime) \rangle = \delta(t-t^\prime)$. Hydrodynamic interactions modify the structure of eq. (\ref{eq2_1}) (or of
eq. (\ref{eq2_3})),  determining the occurrence of memory effects in the
formulation of the equation of motion of a Brownian particle. In the latter
 case, the
thermal force $R(t)$ is no longer $\delta$-correlated,  and its stochastic structure
should be determined (representing the essence of the FD2k and FD3k relations).
Three major classes of problems should be considered:
\begin{enumerate}
\item Memory effects in the dissipative term,
\begin{equation}
\frac{d v}{d t}= - \int_0^t \, h(t-\tau) \,  v(\tau) \, d \tau + R(t)
\label{eq2_4}
\end{equation}
with  $h(t) \geq 0$, $\int_0^\infty h(t) \, dt =1$, such that
the dynamics associated with $h(t)$ is dissipatively stable
\cite{gpp_localrepr}.
The boundedness of the integral of $h(t)$ implies the existence of
a finite non-zero value of the particle diffusivity, i.e. an asymptotic
linear scaling of the particle mean square displacement with time.
 This model  physically corresponds
to the motion in a non-Newtonian linear viscoelastic fluid, in which the Stokes approximation 
for
 the force exerted by the fluid on the Brownian particle is assumed.
Eq. (\ref{eq2_4}) represents the  so called generalized Langevin equation with memory
treated by  Zwanzig  \cite{zwanzig} and Kubo \cite{kubo}. 
\item Memory effects in the fluid inertial contribution,
\begin{equation}
\frac{d v}{d t}= - v - \int_0^t k(t-\tau) \, \frac{d v(\tau)}{d \tau} \, d \tau + R(t)
\label{eq2_5}
\end{equation}
with $k(t) \geq 0$.
Eq. (\ref{eq2_5}) corresponds to the classical model of a Brownian particle in 
a Newtonian fluid in which the fluid inertia is taken into account, and the
mass considered is the extended mass \cite{landau}.  
The force  exerted by the fluid on the particle
stems from solution of  the time-dependent Stokes equation.
\item Memory effects  both in  the dissipative and in the fluid inertial contributions:
\begin{equation}
\frac{d v}{d t}=  - \int_0^t \, h(t-\tau) \,  v(\tau) \, d \tau
- \int_0^t k(t-\tau) \, \frac{d v(\tau)}{d \tau} \, d \tau + R(t)
\label{eq2_6}
\end{equation}
with $h(t)>0$, summable and dissipatively stable and $k(t)>0$.

Equation (\ref{eq2_6}) can be viewed as the generalization of
eqs. (\ref{eq2_4}) and (\ref{eq2_5}) for a linear viscoelastic fluid,
and more generally for any isotropic complex fluid in the linear
time-dependent hydrodynamic regime.
This represents the most complete description of Brownian dynamics in
a unconfined fluid phase.
\end{enumerate}

As addressed in \cite{gionaklages}, eqs. (\ref{eq2_5})-(\ref{eq2_6}) apply
for $v(t=0)=v(0)=0$. In the general case, they should contain
 an impulsive initial condition, so that
eqs. (\ref{eq2_5})-(\ref{eq2_6})  are substituted by
the corresponding equations
\begin{equation}
\frac{d v}{d t}= - v - \int_0^t k(t-\tau) \,  \left (
\frac{d v(\tau)}{d \tau} + v(0) \, \delta(\tau) \right ) \, d \tau + R(t)
\label{eq2_8}
\end{equation}
\begin{equation}
\frac{d v}{d t}=  - \int_0^t \, h(t-\tau) \,  v(\tau) \, d \tau
- \int_0^t k(t-\tau) \, \left ( \frac{d v(\tau)}{d \tau}
+ v(0) \, \delta(\tau) \right )\, d \tau + R(t)
\label{eq2_9}
\end{equation}

\section{Modal representation}
\label{sec3}

Let us consider the case where the memory kernels admit a modal representation
in terms of sums of exponentially decreasing functions of time.
As  discussed in the introduction this represents either
the classical rheological formulation \cite{makosko,rheol}
or the stochastic assumption amenable to a Markovian embedding
of the particle dynamics or the consequence of the principle
of local realizability formulated in \cite{gpp_localrepr}.

In the discrete case, this means that there exist  integers $N_d$,
and $N_i$,  and constant $a_i>0$, $\lambda_i>0$, $\gamma_\alpha>0$,
and $\mu_\alpha>0$, such that,
\begin{equation}
h(t)= \sum_{i=1}^{N_d} a_i \, e^{-\lambda_i \, t} \, , \qquad
k(t) = \sum_{\alpha=1}^ {N_i} \gamma_\alpha \, e^{-\mu_\alpha \, t}
\label{eq3_1}
\end{equation}
where in principle, either $N_d$ or $N_i$, or both,  could be infinite.
Eq. (\ref{eq3_1})  represents a model for the memory response (both
dissipative and inertial) characterized by  linear relaxation
decays (with timescales $\lambda_i^{-1}$ and  $\mu_\alpha^{-1}$, respectively).
 There is no direct elementary relation between the
spectrum of rates associated with the fluid inertial contribution $\{ \mu_\alpha \}_{\alpha=1}^{N_i}$
and the spectrum of relaxation rates $\{\lambda_i \}_{i=1}^{N_d}$
describing viscoelastic dissipation.
In order to maintain the distinction between these two mechanisms, we use latin
indexes for the dissipative terms, and greek ones for those
pertaining to the fluid-inertial contribution.

Eq. (\ref{eq3_1}) can be extended to the case of a  continuous
spectrum of relaxation rates as
\begin{equation}
h(t)= \int_0^\infty a(\lambda) \, e^{-\lambda \, t} \, d \lambda \, ,
\qquad
k(t)= \int_0^\infty \gamma(\mu) \, e^{-\mu \, t} \, d \mu 
\label{eq3_2}
\end{equation}
and therefore $h(t)$ and $k(t)$ can be viewed as the Laplace transforms
of the relaxational dissipative and inertial spectra $a(\lambda)$,
and $\gamma(\mu)$, respectively.

As regards $h(t)$, the existence of a long-term friction factor  $\eta$,
and the normalization considered eq. (\ref{eq2_2}), extended to
the memory case,  imply that
$\int_0^\infty h(t) \, d t=1$, and consequently we have in the discrete/continuous case
\begin{equation}
\sum_{i=1}^{N_d} \frac{a_i}{\lambda_i} = 1 \, ,
\qquad \mbox{or} \qquad
\int_0^\infty \frac{a(\lambda)}{\lambda} \, d \lambda =1
\label{eq3_3}
\end{equation}
In the continuous case, this implies  the asymptotics
$a(\lambda)= O(\lambda^\varepsilon)$, $\varepsilon>0$ for $\lambda \rightarrow
0$, and $a(\lambda)=O(\lambda^{-\delta})$, $\delta>0$ 
for $\lambda \rightarrow \infty$. Observe that eq. (\ref{eq3_3}) does not
represent a physical principle, but solely a normalization condition.
The physical constraint underlying eq. (\ref{eq3_3}) is  that
the quantities $\sum_{i=1}^{N_d} \frac{a_i}{\lambda_i}$, $\int_0^\infty \frac{a(\lambda)}{\lambda} \, d \lambda$ should be bounded, corresponding to the finiteness of the long-term friction factor.
While $a_i>0$,  $i=1,\dots,N_d$, ensures the dissipative stability of the model,
the condition $\mu_\alpha>0$, $\alpha=1,\dots,N_i$,
 are consistent with the structure of the 
generalized Basset kernel in a Maxwell fluids \cite{gionavisco} and
are further discussed in Section \ref{sec5}.

As regards the fluid inertial kernel $k(t)$, we have in the
Newtonian case $k(t) \sim 1/\sqrt{t}$, and therefore $\lim_{t \rightarrow 0}
k(t)= \infty$. However,  as shown in \cite{gionavisco}, it is
a generic property of viscoelastic fluids that 
\begin{equation}
\lim_{t \rightarrow 0} k(t)= k_0 < \infty
\label{eq3_4}
\end{equation}
 and this is physically associated with the bounded velocity
in the propagation of the shear stresses. As a consequence, we have
the general conditions
\begin{equation}
\sum_{\alpha=1}^{N_i} \gamma_\alpha = k_0 < \infty  \qquad \mbox{or} \qquad \int_0^\infty \gamma(\mu) \, d \mu = k_0 < \infty
\label{eq3_5}
\end{equation}
As discussed in \cite{part2}, if $N_d$ or $N_i$ go to infinity,
the long-term scaling of $h(t)$ or $k(t)$, respectively,
may be different from an exponential behaviour, and e.g. may
attain a power-law scaling, which can describe, within
the hydromechanic formalism, the occurrence of anomalous
diffusive properties.

\section{The dissipative case}
\label{sec4}

In the presence of pure dissipation, the
FD2k relation is  subsumed in     the Kubo theorem \cite{kubo,kubolibro}, 
assessing that at thermal equilibrium, the correlation function
of the stochastic force is proportional to the dissipative
memory kernel, i.e., $\langle R(t) \, R(0) \rangle \sim h(t)$.
An explicit representation of $R(t)$ in the modal case has been 
developed by Goychuk \cite{goychuk} (see also \cite{mod1,mod2,mod3}).
In this Section we develop another equivalent approach to this problem,
in order   to introduce
 the moment formalism will be used in the
remainder for addressing the general case.

Consider eqs. (\ref{eq2_4}) and the modal
expansion eq. (\ref{eq3_1}) for $h(t)$,
\begin{equation}
\frac{d v}{d t}= - \sum_{i=1}^{N_d} a_i e^{-\lambda_i  \, t} *
v(t) + R(t)
= - \sum_{i=1}^{N_d} a_i \, e^{-\lambda_i  \, t} * \left ( v(t)+ r_i(t)
\right )
\label{eq4_1}
\end{equation}
where ``$*''$ indicates convolution.
In eq. (\ref{eq4_1}) we have redistributed the fluctuational contribution
amongst the dissipative modes, introducing the modal fluctuation
terms $r_i(t)$, $i=1,\dots,N_d$.
Setting
\begin{equation}
\theta_i(t) = e^{-\lambda_i \, t} * \left (( v(t)+ r_i(t)
\right )
\label{eq4_2}
\end{equation}
eq. (\ref{eq4_1}) can be expressed   as a system of stochastic
differential equations
\begin{eqnarray}
\frac{d v(t)}{d t} & =  &- \sum_{i=1}^{N_d} a_i \, \theta_i(t) \nonumber \\
\frac{d \theta_i(t)}{d t } & = & - \lambda_i \, \theta_i(t) + v(t) + r_i(t)
\label{eq4_3}
\end{eqnarray}
For $r_i(t)$, assume a Wiener representation, i.e.
\begin{equation}
r_i(t)=  \sqrt{2} b_i \, \xi_i(t) \, , \qquad i=1,\dots,N_d
\label{eq4_4}
\end{equation}  
where  $\xi_i(t)$ are the 
distributional derivatives of mutually
independent Wiever processes, thus $\langle
\xi_i(t) \, \xi_j(t^\prime) \rangle =  \delta_{i,j} \, \delta(t-t^\prime)$,
and  the coefficients $b_i$ should be determined from the Langevin condition
eq. (\ref{eq1_2}) that, expressed in terms of the mutually independent
modal fluctuations  $r_i(t)$, takes the form 
\begin{equation}
\langle r_i(t) \, v(0) \rangle_{\rm eq} =0 \, , \qquad i=1,\dots,N_d
\label{eq4_4bis}
\end{equation}
The  normalized velocity autocorrelation function $C_{vv}(t)=\langle v(t) \, v(0) \rangle_{\rm eq}$ is
the solution of the  equation
\begin{equation}
\frac{d C_{vv}(t)}{d t}= - \sum_{i=1}^{N_d} a_i \, e^{-\lambda_i \, t} *
C_{vv}(t) \, , \qquad C_{vv}(0)=1
\label{eq4_5}
\end{equation}
Indicating with
$C_{\theta_i v}(t)= \langle \theta_i(t) \, v(0) \rangle_{\rm eq}$ the correlation
function of $\theta_i(t)$ with $v(0)$ at equilibrium,
it follows from eqs. (\ref{eq4_3}) and (\ref{eq4_4bis}) that
\begin{eqnarray}
\frac{d C_{vv}(t)}{d t} & =  &- \sum_{i=1}^{N_d} a_i \, C_{\theta_i v}(t) \nonumber
\\
\frac{d C_{\theta_i v}(t)}{d t} &=  &- \lambda_i \, C_{\theta_i v}(t) + C_{vv}(t)
\label{eq4_6}
\end{eqnarray}
The latter system of equations
admits the solution $C_{\theta_i v}(t) = C_{\theta_i v}(0) \, e^{-\lambda_i \, t}
+ e^{-\lambda_i \, t} * C_{vv}(t)$, and thus
\begin{equation}
\frac{d C_{vv}(t)}{d t}=
- \sum_{i=1}^{N_d} a_i \, e^{-\lambda_i \, t} * C_{vv}(t) - \sum_{i=1}^{N_d}
a_i \, C_{\theta_i v}(0) \, e^{-\lambda_i \, t}
\label{eq4_7}
\end{equation}
Eq. (\ref{eq4_7}) coincides with eq. (\ref{eq4_5}) provided that
\begin{equation}
C_{\theta_i v}(0)= \langle \theta_i \, v \rangle_{\rm eq} =0 \, , \qquad i=1,\dots,N_d
\label{eq4_8}
\end{equation}
and $C_{vv}(0)=1$, where $\langle \theta_i \, v \rangle_{\rm eq}$ represents the
second-order mixed moment of $\theta_i$ and $v$ at equilibrium.
The expansion coefficients $b_i$, $i=1,\dots,N_d$ can be
thus determined  imposing the $N_d$-relations eq. (\ref{eq4_8}).

To this purpose, consider the Fokker-Planck equation
for the density $p(v,\boldsymbol{\theta},t)$, $\boldsymbol{\theta}=(\theta_1\,\dots,\theta_{N_d})$ associated with eqs. (\ref{eq4_3})-(\ref{eq4_4}),
\begin{eqnarray}
\frac{\partial p}{\partial t} &= & \sum_{h=1}^{N_d} a_h \, \theta_h \, \frac{\partial
p}{\partial v} + \sum_{h=1}^{N_d} \frac{\partial}{\partial \theta_h} \left (
\lambda_h \, \theta_h \, p \right )
- v \sum_{h=1}^{N_d} \frac{\partial p}{\partial \theta_h} 
+  \sum_{h=1}^{N_d} b_h^2 \, \frac{\partial^2 p}{\partial \theta_h^2}
\label{eq4_9}
\end{eqnarray}
from which, the dynamics of the second-order moments,
\[
m_{vv}(t)= \langle v^2(t) \rangle \, , \quad m_{\theta_i v}= \langle \theta_i(t) v(t)
\rangle \, , \quad m_{\theta_i \theta_j}= \langle \theta_i(t) \theta_j(t) \rangle
\]
follows
\begin{eqnarray}
\frac{d m_{vv}}{d t} & =  &-  2 \sum_{h=1}^{N_d} a_h \, m_{\theta_h v}
\nonumber \\
\frac{d m_{\theta_i v}}{d t} & =  & - \sum_{h=1}^{N_d} a_h \,
m_{\theta_i \theta_h} - \lambda_i \, m_{\theta_i v} + m_{vv}
\label{eq4_10} \\
\frac{d m_{\theta_i \theta_j}}{d t} & = & - (\lambda_i + \lambda_j)
\, m_{\theta_i \theta_j}+ m_{\theta_i v} + m_{\theta_j v} +
2 \, b_i^2 \, \delta_{ij}
\nonumber 
\end{eqnarray}
where $\delta_{ij}$ are the Kronecker symbols.
Indicating with $m_{vv}^*=\langle v^2 \rangle_{\rm eq}$  the 
equilibrium (steady-state) value of $m_{vv}$, and similarly for the other moments,
and imposing the conditions
eq. (\ref{eq4_8}), from the third equation (\ref{eq4_10})
one obtains
\begin{equation}
m_{\theta_i \theta_j}^* = \frac{b_i^2}{\lambda_i} \, \delta_{ij}
\label{eq4_11}
\end{equation}
that substituted into the second equation (\ref{eq4_10}), for
$m_{vv}^*=1$,
provides $a_i b_i^2/\lambda_i=1$, i.e.,
\begin{equation}
b_i= \sqrt{\frac{\lambda_i}{a_i}} \, , \qquad i=1,\dots,N_d
\label{eq4_12}
\end{equation}
defining the thermal fluctuations, thus solving FD3k.

In the continuous case, associated with  the spectrum $a(\lambda)$
of dissipation rates, a continuous spectrum
of internal degrees of freedom $\theta_\lambda(t)$, $\lambda \in [0,\infty)$, 
can be defined, 
and each $\theta_\lambda(t)$ satisfies the equation
\begin{equation}
\frac{d \theta_\lambda(t)}{d t}= - \lambda \, \theta_\lambda(t)
+ v(t) + \sqrt{2} \, b(\lambda) \, \xi_\lambda(t)
\label{eq4_13}
\end{equation}
with $\xi_\lambda(t) \, \xi_{\lambda^\prime}(t^\prime) \rangle = 
\delta(\lambda-\lambda^\prime) \, \delta(t-t^\prime)$.
This leads to the continuous analogue of eq. 
(\ref{eq4_12}), namely
\begin{equation}
b(\lambda)= \sqrt{\frac{\lambda}{a(\lambda)}}
\label{eq4_14}
\end{equation}
The overall  stochastic thermal force is thus given by
\begin{equation}
R(t)=  \sqrt{2} \, \int_0^\infty a(\lambda)  \, b(\lambda)
\, e^{-\lambda \, t} * \xi_\lambda(t) \, d \lambda
\label{eq4_15}
\end{equation}
and this result is consistent with Kubo theory since
\begin{equation}
\langle R(t) \, R(t^\prime) \rangle =
2 \, \int_0^\infty \sqrt{a(\lambda) \, \lambda} \, d \lambda
\int_0^t e^{-\lambda \, (t-\tau)} \, d \tau \int_0^\infty
\sqrt{a(\lambda^\prime) \, \lambda^\prime} \, d \lambda^\prime
\int_0^{t^\prime} e^{-\lambda^\prime \, (t^\prime-\tau^\prime)}
\langle \xi_\lambda(\tau) \, \xi_{\lambda^\prime}(\tau^\prime) \rangle
\, d \tau^\prime
\label{eq4_16}
\end{equation}
Setting $t>t^\prime$ without loss of generality, it follows
from eq. (\ref{eq4_16}) after elementary manipulations
\begin{equation}
\langle R(t) \, R(t^\prime) \rangle 
= \int_0^\infty a(\lambda) \, e^{-\lambda (t- t^\prime)} \, d \lambda + \rho(t,t^\prime)
\label{eq4_17}
\end{equation}
where $\rho(t,t^\prime) \rightarrow 0$ for $t, t^\prime \rightarrow \infty$,
consistently with the Kubo result $\langle R(t) \, R(t^\prime) \rangle = h(|t-t^\prime|)$. Numerical examples are discussed in the Appendix \ref{app2}.

\section{Inertial effects: initial conditions and representations}
\label{sec5}

The equations of motion for a particle immersed in a fluid in which fluid inertial
effects are considered in the limit of negligible Reynolds number
 can be derived using integral transforms (Laplace transform)
to express the force   $F(t)=F_{f \rightarrow p}[v(t)]$ 
exerted by the fluid on the particle \cite{kim}.
In the case of a  rigid 
spherical particle of mass $m$ and radius $R_p$ immersed in a Newtonian fluid in
 the time-dependent
Stokes regime,  one obtains for  the Laplace transform $\widehat{F}(s)$ 
of $F(t)$ the expression
\begin{equation}
-\widehat{F}(s)= \eta \, \widehat{v}(s) + \widehat{k}(s) \, s \, \widehat{v}(s) + m_a s \, \widehat{v}(s)
\label{eq6_1}
\end{equation}
where
$\eta=6 \, \pi \, \mu \, R_p$, $\mu$ being the fluid viscosity, $\widehat{k}(s)= 6 \, \pi \, \sqrt{\rho \, \mu}  \, R_p^2/\sqrt{s}$, where $\rho$ is the fluid density, and $m_a=2 \pi R_p^3 \rho/3$  the added mass.
Transforming eq. (\ref{eq6_1}) in time domain, and inserting it into  
the equation of motion of the particle,
one obtains
\begin{equation}
(m+ m_a) \, \frac{d v(t)}{d t}= - \eta  \, v(t) - k(t) * \left ( \frac{d v(t)}{d t}+ v(0) \, \delta(t)
\right )
\label{eq6_2}
\end{equation}
equipped with the initial condition $\lim_{t \rightarrow 0^+}v(t)=m \, v(0)/(m+m_a)$,
where $k(t)= 6 \, \sqrt{\pi \, \rho \, \mu} \, R_p^2/\sqrt{t}$.
The presence of the term $v(0) \, \delta(t)$ in the fluid inertial contribution is fundamental in order to
provide a correct representation of the LRT. 

Next, consider the more
general  nondimensional equation (\ref{eq2_9}). It is convenient
to rewrite this equation in a slightly different way, making use
of the result derived in \cite{gionavisco}, and expressed by
eq. (\ref{eq3_4}),  indicating that, whenever viscoelastic effects
are considered, the value $k(0)$ exists and  is finite.
Integrating by parts, the fluid inertial contribution entering
eq. (\ref{eq2_9}) can be 
expressed as
\begin{eqnarray}
\int_0^t k(t-\tau) \left ( \frac{d v(\tau)}{d \tau} + v(0) \, \delta (\tau)
\right ) \, d \tau\ &= & \int_0^t \frac{\partial }{\partial \tau}
\left ( k(t-\tau) \, v(\tau) \right ) \, d \tau  \nonumber \\
& - & \int_0^t \frac{\partial k(t-\tau)}{\partial \tau} \, v(\tau) \, d \tau
+ k(t) \, v(0) \nonumber \\
& = & k(0) \, v(t) - k(t) \, v(0) - \int_0^t  \frac{\partial k(t-\tau)}{\partial \tau} \, v(\tau) \, d \tau + k(t) \, v(0) 
\nonumber \\
& = & k(0) \, v(t) - \int_0^t \frac{\partial k(t-\tau)}{\partial \tau} \, v(\tau) \, d \tau
\label{eq6_3}
\end{eqnarray}
This  expression will be used in the next section.

\section{Dissipative/inertial memory effects}
\label{sec6}

Consider  particle dynamics in which both  the dissipative term and the 
fluid inertial contribution
account for memory effects. Enforcing the representation developed in the 
previous section eq. (\ref{eq6_3}), and  the modal
decomposition of the kernels $h(t)$, $k(t)$ eq.  (\ref{eq3_1}), we have in 
 nondimensional form
\begin{eqnarray}
\frac{d v(t)}{d t} & = & - \sum_{h=1}^{N_d} a_h \, e^{-\lambda_h \, t} * v(t) - \sum_{\alpha=1}^{N_i}
\gamma_\alpha e^{-\mu_\alpha \, t} * \left ( \frac{d v(t)}{d t}+ v(0) \, \delta(t) \right ) + R(t)
\nonumber \\
& = & - \sum_{\alpha=1}^ {N_i} \gamma_\alpha \, v(t) - \sum_{i=1}^{N_d} a_h \, e^{-\lambda_h \, t} * v(t)
+ \sum_{\alpha=1}^{N_i} \gamma_\alpha \, \mu_\alpha \, e^{-\mu_\alpha \, t} * v(t) + R(t)
\label{eq7_1}
\end{eqnarray}
As  stated in section \ref{sec3}, for the sake of notational clarity,  we
keep latin indexes for referring to variables
associated with the  dissipative modes,  and greek ones for the
inertial modes.
Consider the following internal variables
\begin{equation}
\theta_h(t) = e^{-\lambda_h \, t} * v(t) \, \quad h=1,\dots,N_d \, , \qquad
z_\alpha(t) = e^{-\mu_\alpha \, t} * v(t) \, \quad \alpha=1,\dots,N_i
\label{eq7_2}
\end{equation}
and assume that the stochastic forcing $R(t)$ can be decomposed
into independent contributions $\xi_i(t)$, $\xi_\alpha^\prime(t)$ acting on
the internal degrees of freedom $\theta_h(t)$, $\zeta_\alpha(t)$, respectively,
and 
expressed as  distributional derivatives of independent Wiener processes,
\begin{equation}
\langle \xi_h(t) \, \xi_k(t^\prime) \rangle = \delta_{hk} \, \delta(t-t^\prime) \, , \quad
\langle \xi_\alpha^\prime(t) \, \xi_\beta^\prime(t^\prime) \rangle = \delta_{\alpha\beta} \, \delta(t-t^\prime)
\, , \quad \langle \xi_h(t) \, \xi_\alpha^\prime(t^\prime) \rangle=0
\label{eq7_2bis}
\end{equation}
Due to the representation of  the inertial effects
involving the presence of two terms, the first one,
$- \sum_{\alpha=1}^{N_i} \gamma_\alpha \, v(t)$ directly proportional
to the velocity, and the other one, $\sum_{\alpha=1}^{N_i} \gamma_\alpha \, 
\mu_\alpha \, e^{-\mu_\alpha \, t} * v(t)$
of convolutional nature, 
 a stochastic contribution
 $\sqrt{2} \, \sum_{\alpha=1}^{N_i} d_\alpha \, \xi_\alpha^\prime(t)$
has been added into the  evolution equation for the velocity, where
the expansion coefficients $d_\alpha$ need to be determined.
Therefore, eq. (\ref{eq7_1}) is recast into a a system of $N_d+N_i+1$ 
differential equations,
\begin{eqnarray}
\frac{d v}{d t} & =  & -  \left (\sum_{\alpha=1}^{N_i} \gamma_\alpha \right ) \, v - \sum_{h=1}^{N_d} a_h \, \theta_h
+  \sum_{\alpha=1}^{N_i} \gamma_\alpha \, \mu_\alpha \, z_\alpha + \sqrt{2} \, \sum_{\alpha=1}^{N_i} d_\alpha \, \xi_\alpha^\prime(t) \nonumber \\
\frac{d \theta_h}{d t} & =  &  - \lambda_h \, \theta_h + v + \sqrt{2} \, b_h \, \xi_h(t)
\label{eq7_3} \\  
\frac{d z_\alpha}{d t}  &=  &- \mu_\alpha \, z_\alpha + v + \sqrt{2} \, c_\alpha \, \xi_\alpha^\prime(t)
\nonumber
\end{eqnarray}
where the $(2 N_i + N_d)$ coefficients, $b_h$ $h=1,\dots,N_d$, $c_\alpha$, $d_\alpha$ $\alpha=1,\dots, N_i$ should
be determined by imposing the conditions 
\begin{equation}
\langle \theta_h \, v \rangle_{\rm eq}= 0 \, ,  \; \; h=1,\dots,N_d  \, , \quad
 \langle z_\alpha \, v \rangle_{\rm eq} =0 \, , \; \; \alpha=1,\dots,N_\alpha
\label{eq7_3bis}
\end{equation}
The Fokker-Planck equation associated with eq. (\ref{eq7_3}) is given by
\begin{eqnarray}
\frac{\partial p}{\partial t} & = &  \left ( \sum_{\alpha=1}^{N_i} \gamma_\alpha \right ) 
  \frac{\partial }{\partial v} \left ( v \, p \right )  +
\left ( \sum_{h=1}^{N_d} a_h   \theta_h  \right ) \, \frac{\partial p}{\partial v} -  \left ( \sum_{\alpha=1}^{N_i}
 \gamma_\alpha \, \mu_\alpha \, z_\alpha \right )
 \frac{\partial p}{\partial v} +  \left ( \sum_{\alpha=1}^{N_i} d_\alpha^2  \right ) 
\, \frac{\partial^2 p}{\partial v^2} \nonumber \\
& + & \sum_{h=1}^{N_d}  \lambda_h \, \frac{\partial }{\partial \theta_h} \left ( \theta_h \, p \right ) - 
v  \, \sum_{h=1}^{N_d} \frac{\partial p}{\partial \theta_h}+ \sum_{h=1}^{N_d} b_h^2 \, 
\frac{\partial^2 p}{\partial  \theta_h^2} + \sum_{\alpha=1}^{N_i} \mu_\alpha \, \frac{\partial }{\partial z_\alpha} \left ( z_\alpha \, p \right )  \label{eq7_4} \\
& - & v \, \sum_{\alpha=1}^{N_i} \frac{\partial p}{\partial z_\alpha}+
2 \sum_{\alpha=1}^{N_i} d_\alpha \, c_\alpha \, \frac{\partial^2 p}{\partial v \partial z_\alpha}
+ \sum_{\alpha=1}^{N_i} c_\alpha^2 \, \frac{\partial^2 p}{\partial z_\alpha^2}
\nonumber 
\end{eqnarray}
from which the evolution equations for the second-order moments follow
\begin{eqnarray}
\frac{d m_{vv}}{d t} &= & - 2 \, \left ( \sum_{\beta=1}^{N_i} \gamma_\beta \right )   \, m_{vv} -
2 \sum_{h=1}^{N_d} a_h \, m_{\theta_h v} + 2 \sum_{\alpha=1}^{N_i} \gamma_\alpha \, \mu_\alpha \, m_{z_\alpha v }
+ 2 \sum_{\alpha=1}^{N_i} d_\alpha^2 \nonumber \\
\frac{d m_{\theta_h v}}{d t} &= &  -  \left ( \sum_{\beta=1}^{N_i} \gamma_\beta \right )  \, 
m_{\theta_h v } - \sum_{k=1}^{N_d} a_k m_{\theta_h \theta_k} + \sum_{\alpha=1}^{N_i} \gamma_\alpha \, \mu_\alpha \, m_{\theta_h z_\alpha} - \lambda_h m_{\theta_h v} + m_{vv} \nonumber \\
\frac  {d m_{z_\alpha v}}{d t}  & = & - \left ( \sum_{\beta=1}^{N_i} \gamma_\beta \right )
\, m_{z_\alpha v} - \sum_{k=1}^{N_i} a_k \, m_{\theta_k z_\alpha} +
\sum_{\beta=1}^{N_i} \gamma_\beta \, \mu_\beta \, m_{z_\alpha z_\beta} - \mu_\alpha \, m_{z_\alpha v} + m_{vv}+ 2 \, d_\alpha c_\alpha \nonumber \\
\frac {d m_{\theta_h \theta_k}}{dt} & = & - (\lambda_h + \lambda_k) \, m_{\theta_h \theta_k} +
m_{\theta_h v} + m_{\theta_k v}+ 2 \, b_h^2 \, \delta_{h,k}
\label{eq7_5} \\
\frac{d m_{\theta_h z_\alpha}}{d t} &= & -\lambda_h \, m_{\theta_h z_\alpha} + m_{z_\alpha v}
- \mu_\alpha \, m_{\theta_h z_\alpha} + m_{\theta_h v} \nonumber  \\
\frac{d m_{z_\alpha z_\beta}}{d t}  & =  & - (\mu_\alpha + \mu_\beta ) \, m_{z_\alpha z_\beta} + m_{z_\alpha v} 
+ m_{z_\beta v} + 2 \, c_\alpha^2 \, \delta_{\alpha , \beta} \nonumber
\end{eqnarray}
The fluctuation-dissipation representation of the stochastic
contribution can be derived  imposing eqs. (\ref{eq7_3bis}),
which imply for the steady-state  values of the mixed
moments the conditions
\begin{equation}
m_{\theta_h v}^*=0 \, , \quad h=1,\dots,N_d \,, \qquad m_{z_\alpha v}^*=0 \, , \quad \alpha=1,\dots,N_i
\label{eq7_6}
\end{equation}
and $m_{vv}^*=1$.
Inserting these conditions into the systems of evolution equation for the second-order moments,
one obtains at steady state the following relations: (i)
from the first equation (\ref{eq7_5})
\begin{equation}
\sum_{\alpha=1}^{N_i} d^2_\alpha = \sum_{\alpha=1}^{N_i} \gamma_\alpha
\label{eq7_7}
\end{equation}
(ii) from the fourth equation (\ref{eq7_5})
\begin{equation}
m_{\theta_h \theta_k}^*= \frac{b_h^2}{\lambda_h} \, \delta_{hk}
\label{eq7_8}
\end{equation}
(iii) from the sixth equation (\ref{eq7_5})
\begin{equation}
m_{z_\alpha z_\beta}^*= \frac{c^2_\alpha}{\mu_\alpha} \, \delta_{\alpha\beta}
\label{eq7_9}
\end{equation}
Moreover, (iv) from the fifth equation (\ref{eq7_5}) we have
\begin{equation}
m_{\theta_h z_\alpha}^*=0
\label{eq7_10}
\end{equation}
Enforcing the relations (\ref{eq7_7})-(\ref{eq7_10}) into the second equation (\ref{eq7_5})  the coefficients $b_h$ are determined
\begin{equation}
b_h^2 = \frac{\lambda_h}{a_h}
\label{eq7_11}
\end{equation}
The third equation (\ref{eq7_5}) provides
\begin{equation}
\sum_{\beta=1}^{N_i} \gamma_\beta \mu_\beta m_{z_\alpha z_\beta}^*+ 1  + 2 \, d_\alpha \, c_\alpha =0
\label{eq7_12}
\end{equation}
which, due to eq. (\ref{eq7_9}), simplifies as
\begin{equation}
\gamma_\alpha \, c_\alpha^2 +1 + 2 \, d_\alpha \, c_\alpha = 0
\label{eq7_13}
\end{equation}
Eq. (\ref{eq7_7}) is certainly satisfied if $d_\alpha$ are related
to $\gamma_\alpha$ in the form
\begin{equation}
d_\alpha= \sqrt{\gamma_\alpha}
\label{eq7_14}
\end{equation}
which, inserted into eq. (\ref{eq7_13}), provides 
\begin{equation}
\gamma_\alpha \, c_\alpha^2 +1 + 2 \, \sqrt{\gamma_\alpha} \, c_\alpha=0
\label{eq7_15}
\end{equation}
admitting the solution $c_\alpha=- 1/\sqrt{\gamma_\alpha}$. Therefore, 
the whole system of   coefficients accounting
for the stochastic fluctuations is given by
\begin{equation}
b_h= \sqrt{\frac{\lambda_h}{a_h}} \, , \quad h=1,\dots,N_d\,, \qquad d_\alpha=\sqrt{\gamma_\alpha} \, , \quad c_\alpha= - \frac{1}{\sqrt{\gamma_\alpha}} \, , \quad \alpha=1,\dots,N_i
\label{eq7_16}
\end{equation}
and this represents a general solution for the representation of stochastic fluctuations, i.e. for  the FD3k  problem,
in  the presence of both dissipative and inertial memory effects.
Observe that the value of the coefficients $b_h$ associated
with the dissipative degrees of freedom concides with eq. (\ref{eq4_12}), i.e.
it is completely independent of the  presence of the fluid inertial term.
Numerical cases studies are addressed in the Appendix B.

A final remark concerns the uniqueness of eq. (\ref{eq7_14}) starting 
from eq. (\ref{eq7_7})
and the positivity of the $\gamma_\alpha$'s.
Consider eq. (\ref{eq7_13}) for $c_\alpha$, with $d_\alpha$ still to be determined.
It admits the solutions
\begin{equation}
c_\alpha= -\frac{d_\alpha}{\gamma_\alpha} \pm \sqrt{ \frac{d_\alpha^2}{\gamma_\alpha^2} - \frac{1}{\gamma_\alpha}}
\label{eq7x1}
\end{equation}
and $c_\alpha$ is real-valued provided that $d_\alpha \geq \gamma_\alpha$.
If for some $\alpha$, $\gamma_\alpha <0$, there is no solution of the FD3k problem
associated with the dynamic scheme eq. (\ref{eq7_1}).
If all the $\gamma_\alpha>0$, any choice $d_\alpha^2>\gamma_\alpha$
violates eq. (\ref{eq7_7}). Therefore the condition eq. (\ref{eq7_14})
is not only sufficient but also necessary, modulo
the fact that $d_\alpha$ and $c_\alpha$ can flip  sign,
i.e. $d_\alpha=-\sqrt{\gamma_\alpha}$,
$c_\alpha=1/\sqrt{\gamma_\alpha}$, $\alpha=1,\dots,N_i$, defining 
a stochastic dynamics fully equivalent to  eq. (\ref{eq7_16}).

\section{Representation of the stochastic force $R(t)$}
\label{sec7}

Consider the general setting eq. (\ref{eq2_9}) with the kernels given by eq. (\ref{eq3_1}),
and the  representation of the dynamics via the internal degrees of freedom
$\theta_h(t)$ and $z_\alpha(t)$ as developed in section \ref{sec6}, where
\begin{equation}
\left \{
\begin{array}{l}
\theta_h(t)  = e^{-\lambda_h \, t} * v(t) + \sqrt{2} \, b_h \, e^{-\lambda_h \, t} * \xi_h(t) \\
z_\alpha(t) = e^{-\mu_\alpha \, t} * v(t) + \sqrt{2} \, c_\alpha \, e^{-\mu_\alpha \, t} * \xi_\alpha^\prime(t)
\end{array}
\right .
\label{eq9_1}
\end{equation}
From eq. (\ref{eq7_3}) the random force $R(t)$ assumes the following representation
\begin{eqnarray}
R(t)& = & \underbrace{- \sqrt{2} \, \sum_{h=1}^{N_d} \, a_h \, b_h \, e^{-\lambda_h \, t } * \xi_h(t)}_{\small dissipative  \; fluctuations} \nonumber \\
& + &  \underbrace{ \sqrt{2} \,  \sum_{\alpha=1}^{N_i}  d_\alpha \, \xi_\alpha^\prime (t) +
\sum_{\alpha=1}^{N_i} \gamma_\alpha \, \mu_\alpha \, c_\alpha e^{-\mu_\alpha \, t} *
\xi_\alpha^\prime(t) }_{\small fluid \; inertial \; fluctuations}  
\label{eq9_2} \\
& = & R_d(t)+ R_i(t) \nonumber
\end{eqnarray}
Substituting the values derived for
 the expansion coefficients eq. (\ref{eq7_16}), these two contributions
take the form
\begin{eqnarray}
R_d(t) & = & - \sqrt{2} \, \sum_{h=1}^{N_d} \sqrt{a_h \, \lambda_h} \, e^{-\lambda_h \, t} * \xi_h(t)
\nonumber \\
R_i(t) & = & \sqrt{2} \,  \sum_{\alpha=1}^{N_i} \sqrt{\gamma_\alpha} \left (
- \mu_\alpha e^{-\mu_\alpha t} * \xi_\alpha^\prime(t) + \xi_\alpha^\prime(t) \right )
\label{eq9_3}
\end{eqnarray}
where $\langle R_d(t) \rangle=\langle R_i(t) \rangle=0$.
By definition of the stochastic forcings
 $\{ \xi_h(t) \}_{h=1}^{N_d}$ and $\{ \xi_\alpha \}_{\alpha=1}^{N_i}$, $R_d(t)$ and $R_i(t)$ are independent of each other, and therefore
\begin{equation}
\langle R_d(t) \, R_i(0) \rangle = \langle R_i(t) \, R_d(0) \rangle =0
\label{eq9_4}
\end{equation}
for any $t$, and $\langle R_d(t) R_d(0) \rangle =h(t)$, as previously derived, consistently with the
Kubo theory.
In order to determine the autocorrelation function of $R_i(t)$, consider the
 process $\xi(t)$ (i.e., the  distributional derivative of a Wiener
process),  its first-order mollification $y(t)=e^{-\mu \, t} * \xi(t)$, $\mu>0$
and the elementary properties
\begin{eqnarray}
\langle \xi(t) \, \xi(0) \rangle & = & \delta(t) \nonumber \\
\langle y(t) \, y(0) \rangle & = & \frac{e^{-\mu \, t}}{2 \, \mu} \nonumber \\
\langle \xi(t) \, y(0) \rangle & = & 0
\label{eq9_5} \\
\langle y(t) \, \xi(0) \rangle & = & e^{-\mu \, t} \nonumber
\end{eqnarray}
Since $R_i(t)$ is the superposition of such processes for different values of the
exponent $\mu=\mu_\alpha$, and these processes,
i.e. $y_\alpha^\prime= e^{-\mu_\alpha \, t} * \xi_\alpha^\prime(t)$,
are independent of each other, it follows that
\begin{eqnarray}
\langle R_i(t) \, R_i(0) & =  & \left \langle 2 \, \sum_{\alpha=1}^{N_i} \sum_{\beta=1}^{N_i}
\sqrt{\gamma_\alpha \, \gamma_\beta} \, \left [- \mu_\alpha \, y_\alpha^\prime(t) + \xi_\alpha^\prime(t)
\right ] \,  \left [-\mu_\beta \, y_\beta^\prime(t) + \xi_\beta^\prime (t) \right ] \right \rangle
\nonumber \\
& = & 2 \, \sum_{\alpha=1}^{N_i} \gamma_\alpha \, \left [ \mu_\alpha^2 \, \langle y_\alpha^\prime(t) \, y_\alpha^\prime(0) \rangle - \mu_\alpha \, \langle y_\alpha^\prime(t) \, \xi_\alpha^\prime(0) \rangle
- \mu_\alpha \, \langle \xi_\alpha(t) \, y_\alpha(0) \rangle + \langle \xi_\alpha(t) \, \xi_\alpha(0) \rangle \right ] \nonumber \\
& = & 2 \, \sum_{\alpha=1}^{N_i} \gamma_\alpha \, \left [ \frac{\mu_\alpha}{2} \, e^{-\mu_\alpha \, t} - \mu_\alpha \, e^{-\mu_\alpha \, t} + \delta(t) \right ] \nonumber \\
& = & - \sum_{\alpha=1}^{N_i} \gamma_\alpha \, \mu_\alpha \, e^{- \mu_\alpha \, t} + 2 \, \sum_{\alpha=1}^{N_i}  \gamma_\alpha \, \delta (t)
\label{eq9_6}
\end{eqnarray}
The latter result can be  expressed in a meaningful way with respect to the fluid inertial
kernel $k(t)$ as
\begin{equation}
\langle R_i(t) \, R_i(0) \rangle  = \frac{d k(t)}{d t} + 2 \, k(0) \, \delta (t)
\label{eq9_7}
\end{equation}
that provides a compact representation of this autocorrelation function and
the solution of FD2k as regards fluid inertial effects.
Including also the dissipative effects,
we arrive to the general formulation of FD2k in particle hydromechanics
\begin{equation}
\langle R(t) \, R(0) \rangle = h(t)+ 2 \, k(0) \, \delta(t) + \frac{d k(t)}{d t}
\label{eq9_7a}
\end{equation}
Observe the importance of the condition   eq. (\ref{eq3_4}), i.e. $k(0) < \infty$,
otherwise  the second term at the r.h.s of eqs. (\ref{eq9_7}) and (\ref{eq9_7a})
 would be meaningless, 
consistently with the analysis
developed in \cite{gionavisco}.

As $k(0) >0$, fluid inertial effects induce necessarily the
presence of a white-noise contribution, the term $2 \, k(0) \, \delta(t)$
at the r.h.s. of eq. (\ref{eq9_7a}) in the autocorrelation
function of $R(t)$. This  term plays an important role 
as regards the regularity properties (and specifically
the H\"older continuity) of the velocity fluctuations
as extensively addressed in \cite{part3}.

\begin{figure}
\includegraphics[width=12cm]{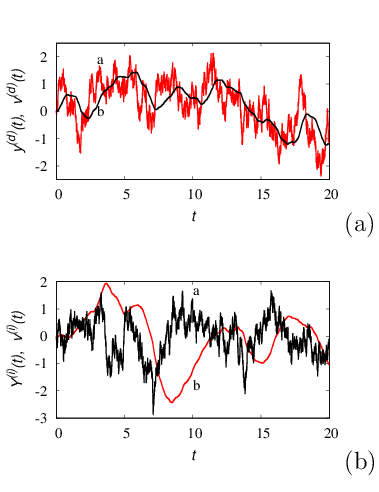}
\caption{Graph of a realization of the elementary stochastic processes associated
with dissipative and fluid inertial memory fluctuations ($\lambda=\mu=1$).
Panel (a): dissipative case: line (a) refers to $y^{(d)}(t)$, line (b) to $v^{(d)}(t)$.
Panel (b): fluid inertial case: line (a) refers to $Y^{(i)}(t)$, line (b) to $v^{(i)}(t)$.}
\label{Fig2fd}
\end{figure}

It follows from the structure of $R_d(t)$ and $R_i(t)$, that the
fluctuations acting on a particle due to the fluid hydrodynamics involve two classes of 
elementary stochastic processes:
\begin{equation}
y^{(d)}(t)= \sqrt{\lambda} \, e^{-\lambda \, t} * \xi(t)
\label{eq9_8}
\end{equation}
associated with the dissipative memory contributions, and 
\begin{equation}
y^{(i)}(t) = \xi^\prime(t) - \mu \, e^{-\mu \, t} * \xi^\prime(t)
\label{eq9_9}
\end{equation}
associated with fluid inertial effects. Let $v^{(d)}(t)$, and $v^{(i)}(t)$ be  the
resultant velocities determined by the sole action of these processes, respectively,
 i.e.,
\begin{eqnarray}
\frac{d v^{(d)}(t)}{d t} & = & - e^{-\lambda \, t} * v^{(d)}(t) + y^{(d)}(t) \nonumber \\
\frac{d v^{(i)}(t)}{d t} & = & - v^{(i)}(t)+ \mu \, e^{-\mu \, t} * v^{(i)}(t) + y^{(i)}(t)
\label{eq9_10}
\end{eqnarray}
assuming $v^{(d)}(0)=v^{(i)}(0)=0$. Figure \ref{Fig2fd} 
 depicts a realization of these
processes:
panel (a) for the dissipative
and  panel (b) for the  fluid inertial case.
 Due to the singularity of $y^{(i)}(t)$, containing a contribution equal to
the distributional derivative of a Wiener process, the integral process $Y^{(i)}(t)=
\int_0^t y^{(i)}(\tau) \, d \tau$ is depicted instead of $y^{(i)}(t)$.

Some qualitative observations emerge from these graphs. In the purely
dissipative case $v^{(d)}(t)$ is simply a mollification of the stochastic forcing
term $y^{(d)}(t)$ corresponding to the dissipative action of a viscoelastically relaxing mode.
Qualitatively different is the behaviour of $v^{(i)}(t)$ with respect to $Y^{(i)}(t)$.
From what derived above, for $t>0$, $\langle R_i(t) \, R_i(0) \rangle <0$, i.e.,
the fluid inertial noise contribution is negatively self-correlated, and also
negatively correlated with respect to  $v^{(i)}(t)$. This
is not surprising, as this term represents the inertial action of nearby fluid elements
opposing to the motion of the particle. Consistently, the
evolution of $Y^{(i)}(t)$ and $v^{(i)}(t)$ is essentially out-of-phase or better to say
in antiphase.

\section{Fluctuational patterns}
\label{sec8}

The modal representations of FD3k obtained in the previous paragraphs  requires further
analysis.  To begin with, it clearly indicates the conceptual and thermodynamic
difference of dissipative and fluid inertial contributions that
 induce structurally different
representations of the action of the thermal fluctuations. Moreover, 
the modal representation of the memory kernels eq. (\ref{eq3_1}) imply the mutual independence of 
the elementary stochastic forcings  acting on each mode.
This suggest the definition of a hierachy of basic fluctuational patterns of increasing
hydrodynamic complexity.
Owing to the linearity of the interactions, and to the
independence of  the stochastic forcings acting on each internal
variable, these patterns are susceptible of composition rules, i.e. of their
own pattern algebra.

For the case of clarity and generality, let us develop the concept of
fluctuational patterns in  a vectorial formulation (as addressed in Appendix \ref{app1}),
for a generic dimension $n=1,2,3,\dots$ of the physical space and in dimensional terms.

First of all, any pattern is defined by: 1)  the dimension $n$ of the space,
2)  the particle mass $m$, 3)  the constant temperature $T$ of
the fluid.  For notational reasons (in order to encompass the nondimensional formulation  as
a particular case), the reference to the constant temperature will be indicated via
the basic thermal energy unit $E=k_B \, T$; 4)  the nature of the hydromechanic
interaction; 5)  the number of independent stochastic forcings involved.
Each basic pattern associated with specific fluid-particle interactions is characterized by
specific hydromechanic parameters and is expressed via a suitable set of internal
degrees of freedom.

Let ${\bf h}(t)$ and ${\bf k}(t)$ be the tensorial dissipative and fluid-inertial memory kernels.
In the dimensional 
formulation of the basic patterns we will adopt the following notation
\begin{eqnarray}
{\bf h}(t) & = &  \sum_{h=1}^{N_d} {\bf a}_h \, \lambda_h \, e^{-\lambda_h \, t} \, ,
\qquad \sum_{h=1}^{N_d} {\bf a}_h = \boldsymbol{\eta}_\infty \nonumber \\
{\bf k}(t) & = & \sum_{\alpha=1}^{N_i} {\bf g}_\alpha \, e^{-\mu_\alpha \, t}
\label{eqpp1}
\end{eqnarray}
so that
$\int_0^\infty {\bf h}(t) \, d t= \boldsymbol{\eta}_\infty$ corresponds to
the long-term friction tensor. Each rate $\lambda_h, \, \mu_\alpha >0$ is real,
and  each hydromechanic matrix, $ {\bf a}_h, {\bf g}_\alpha$,
is symmetric and positive definite.  The nondimensional case follows by setting $m=E=1$.

Each pattern is  indicated by a code symbol and by the fundamental
information on $(m, \, E,\, n,\,N_d,\, N_i)$ where $N_i$ and $N_d$ are,
as in the previous Sections  the  number of independent stochastic forcings
associated with dissipative and fluid-inertial degree of freedom, and
a superscript $(n,N_d,N_i)$ is attached to each pattern symbol.
Moreover, all the different stochastic forcings are independent of each other.

The simplest case is represented by the pattern ${\mathbb S}_{m,E}^{(n,1,0)}(\boldsymbol{\xi}(t))$,
corresponding to the fluid-particle interaction in the instantaneous Stokes regime (${\mathbb S}$ as Stokes),
\begin{equation}
m \, \dot{\bf v}(t) = - \boldsymbol{\eta} \, {\bf v}(t) + \sqrt{2} \, {\bf a} \, \boldsymbol{\xi}(t)
\label{eqpp2}
\end{equation}
where $\dot{\bf v} $ stands for for time derivative $d {\bf v}(t)/dt$.
Observe that here, as well as in the remainder $\boldsymbol{\xi}(t)$ represents the derivative
of an $n$-dimensional  Wiener process. In this case the fluctuational
parameter is ${\bf a}= \sqrt{k_B \, T} \, \boldsymbol{\eta}^{1/2}$.

Next consider the case of a purely dissipative memory dynamics,  with no fluid-inertial
effects, the basic pattern of which is ${\mathbb D}_{m,E}^{(n,N_d,0)}(\{ \boldsymbol{\xi}_h^{(d)}(t) \}_{h=1}^{N_d})$, (${\mathbb D}$ as dissipative),
defined by the dynamics
\begin{eqnarray}
m \, \dot{\bf v}(t) & = & -  \sum_{h=1}^{N_d} {\bf a}_h \, \boldsymbol{\theta}_h(t)
\nonumber \\
\dot{\boldsymbol{\theta}}_h(t)  & = & - \lambda_h \, \boldsymbol{\theta}_h(t) + \lambda_h \, {\bf v}(t) +
\sqrt{2} \, {\bf b}_h \, \lambda_h \, \boldsymbol{\xi}_h^{(d)}(t) \, , \qquad h=1,\dots,N_d
\label{eqpp3}
\end{eqnarray}
where
\begin{equation}
{\bf b}_h = \sqrt{k_B \, T} \,  {\bf a}_h^{-1/2} \, , \qquad h=1,	\dots, N_d
\label{eqpp4}
\end{equation}
The non-dimensional formulation eq. (\ref{eq4_12}), here indicated with the superscript ``(nd)``,
 is recovered by setting $m=E=1$, since from the comparison between eqs. (\ref{eq4_3})-(\ref{eq4_4})
 and eq. (\ref{eqpp3}) we have ${\bf b}_h^{({\rm nd})}= {\bf b}_h$, ${\bf a}_h^{({\rm nd})}=
 {\bf a}_h \, \lambda_h$, and thus, from eq. (\ref{eqpp4}) we
obtain
${\bf b}_h^{({\rm nd})}= \sqrt{\lambda_h} \, \left ({\bf a}_h^{({\rm nd})} \right )^{-1/2}$,
consistently with the scalar relation eq.  (\ref{eq4_12}) and with the vectorial formulation
eq. (\ref{eq8_15}), where $\boldsymbol{\sigma}^{(h)}$ coincides with ${\bf b}_h^{({\rm nd})}$.

\begin{table}
\begin{tabular}{c|c|c|c|c}
\hline
 & & & & \\
$\;\;$ Pattern   $\;\;$  &  $\;\;$   Hydromechanic  par.  $\;\;$   & $\;\;$ Fluctuational par. $\;\;$ & $\;\;$ Auxiliary var. $\;\;$ & $\;\;$ Stochastic forc. $\;\;$ \\
 & & & & \\
\hline
    & & & & \\
${\mathbb S}_{m,E}^{(n,1,0)}$ & $\boldsymbol{\eta}$ &  ${\bf a}= \sqrt{k_B \, T} \,  \boldsymbol{\eta}^{1/2}$& - & $\boldsymbol{\xi}(t)$ \\
& & & &  \\
\hline
& & & &  \\
 ${\mathbb D}_{m,E}^{(n,N_d,0)}$ &   $\{ {\bf a}_h \}_{h=1}^{N_d}$,  $\{ \lambda_h \}_{h=1}^{N_d}$ &  $\{ {\bf b}_h \}_{h=1}^{N_d}$ &  $\{ \boldsymbol{\theta}_h \}_{h=1}^{N_d}$&   $\{ \boldsymbol{\xi}^{(d)}_h(t) \}_{h=1}^{N_d}$ \\
& & ${\bf b}_h=\sqrt{k_B \, T} \, {\bf a}_h^{-1/2}$ & &  \\
\hline
& & {\bf a},  $\{ {\bf c}_\alpha \}_{\alpha=1}^{N_i}$,  $\{ {\bf d}_\alpha \}_{\alpha=1}^{N_i}$ & &  \\
 ${\mathbb I}_{m,E}^{(n,1,N_i)}$ &    $\boldsymbol{\eta}$, $\{ {\bf g}_\alpha \}_{\alpha=1}^{N_i}$,  $\{ \mu_\alpha \}_{\alpha=1}^{N_i}$  &  ${\bf c}_\alpha= - \sqrt{k_B \, T} \, {\bf g}_\alpha^{-1/2}$ &$\{ {\bf z}_\alpha \}_{\alpha=1}^{N_i}$&   $\boldsymbol{\xi}(t)$, $\{ \boldsymbol{\xi}^{(i)}_\alpha(t) \}_{\alpha=1}^{N_i}$ \\
& & ${\bf d}_\alpha= \sqrt{k_B \, T} \, {\bf g}_\alpha^{1/2}$ & &  \\
\hline
&    $\{ {\bf a}_h \}_{h=1}^{N_d}$,  $\{ \lambda_h \}_{h=1}^{N_d}$  &  $\{ {\bf b}_h \}_{h=1}^{N_d}$ &  & \\
${\mathbb D}{\mathbb I}_{m,E}^{(n,N_d,N_i)}$ &  $\{ {\bf g}_\alpha \}_{\alpha=1}^{N_i}$,  $\{ \mu_\alpha \}_{\alpha=1}^{N_i}$ &  $\{ {\bf c}_\alpha \}_{\alpha=1}^{N_i}$,  $\{ {\bf d}_\alpha \}_{\alpha=1}^{N_i}$  &  $\{ \boldsymbol{\theta}_h \}_{h=1}^{N_d}$, $\{ {\bf z}_\alpha \}_{\alpha=1}^{N_i}$ &   $\{ \boldsymbol{\xi}^{(d)}_h(t) \}_{h=1}^{N_d}$, $\{ \boldsymbol{\xi}^{(i)}_\alpha(t) \}_{\alpha=1}^{N_i}$ \\
& & & &  \\
\hline
\end{tabular}
\caption{Synoptic review of the fluctuational patterns, as regards hydromechanic and fluctuational parameters, auxiliary variables and stochastic forcings. $E=k_B \, T$, by definition.}
\label{tab1}
\end{table}

\begin{figure}
    \includegraphics[width=1\linewidth]{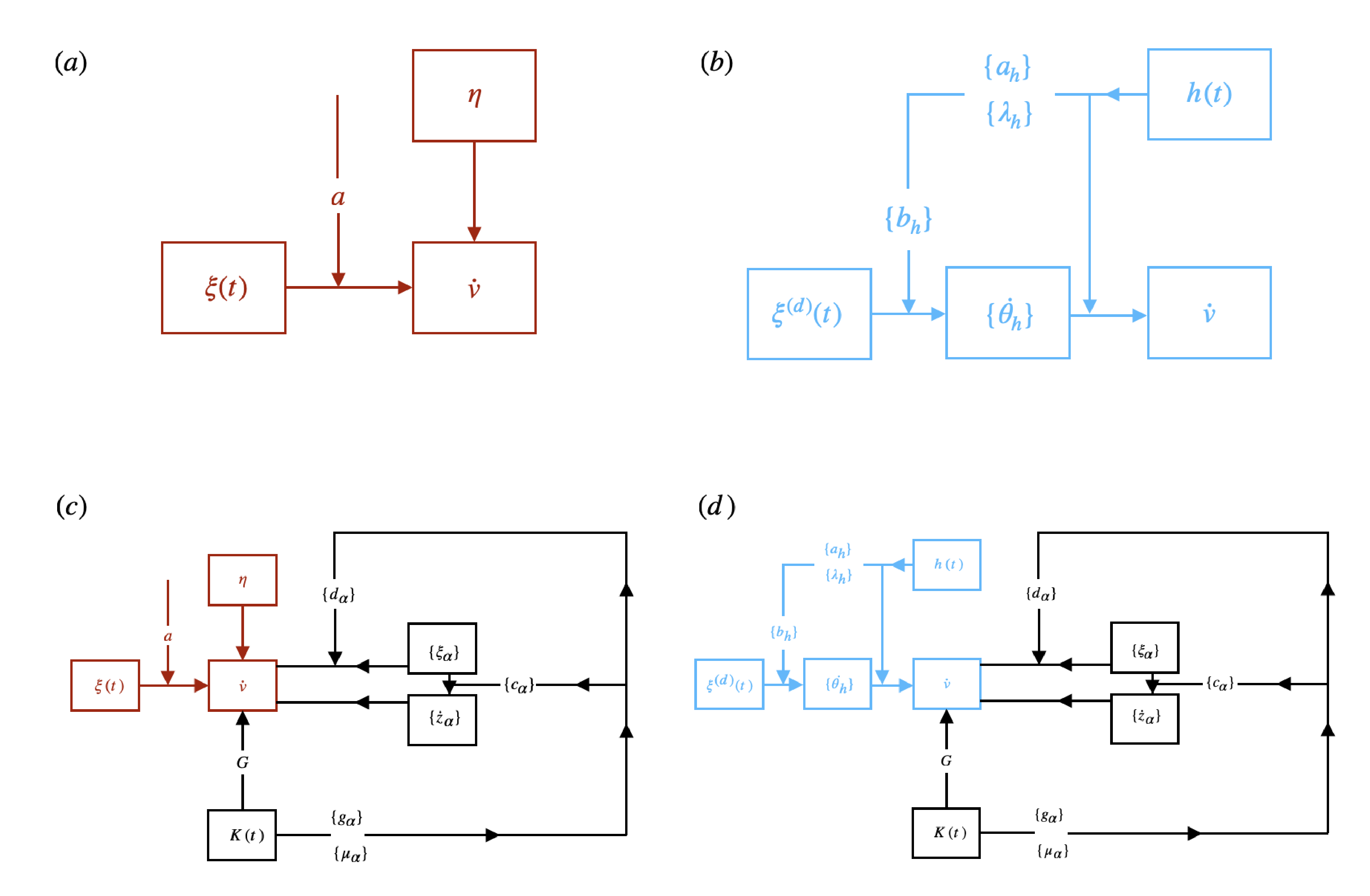}
    \caption{Block diagram of the representation of the  basic fluctuational patterns: (a)
${\mathbb S}_{m,E}^{(n,1,0)}$, (b) ${\mathbb D}_{m,E}^{(n,N_d,0)}$, (c) ${\mathbb I}_{m,E}^{(n,1,N_i)}$, (d) ${\mathbb D}{\mathbb I}_{m,E}^{(n,N_d,N_i)}$.}
    \label{figblock}
\end{figure}

The other basic fluctuational pattern ${\mathbb I}_{m,E}^{(n,1,N_i)}(\boldsymbol{\xi}(t),
\{ \boldsymbol{\xi}_\alpha^{(i)}(t) \}_{\alpha=1}^{N_i})$, (${\mathbb I}$ as Inertial)
involves the fluid-inertial memory dynamics in the presence of a Stokesian
dissipative contribution. Apart from  a very peculiar and anomalous case study of fluid-inertia
acting in the absence of dissipation addressed in part II, a non-vanishing dissipative term ensures 
the existence and achievement of  equilibrium conditions.
The dimensional dynamic equations of this pattern read as
\begin{eqnarray}
m \, \dot{\bf v}(t) & = & - \boldsymbol{\eta} \, {\bf v}(t) - {\bf G} \, {\bf v}(t) +
\sum_{\alpha=1}^{N_i} {\bf g}_\alpha \, \mu_\alpha \, {\bf z}_\alpha(t) +
\sqrt{2} \, {\bf a} \, \boldsymbol{\xi}(t) +
\sqrt{2} \sum_{\alpha=1}^{N_i} {\bf d}_\alpha \, \boldsymbol{\xi}_\alpha^{(i)}(t) \nonumber \\
\dot{\bf z}_\alpha(t)& = & - \mu_\alpha \, {\bf z}_\alpha(t) + {\bf v}(t) + \sqrt{2} \, {\bf c}_\alpha
\, \boldsymbol{\xi}_\alpha^{(i)}(t) \, , \qquad \alpha=1,\dots,N_i
\label{eqpp5}
\end{eqnarray}
where ${\bf G}=\sum_{\alpha=1}^{N_i} {\bf g}_\alpha$,  the fluctuation
coefficient ${\bf a}$ coincides with that of the Stokesian pattern, ${\bf a}=\sqrt{k_B \, T} \, \boldsymbol{\eta}^ {1/2}$, and
\begin{equation}
{\bf  d}_\alpha= \sqrt{k_B \, T} \, {\bf g}_\alpha^{1/2} \, , \quad
{\bf c}_\alpha= - \sqrt{k_B \, T} \, {\bf g}_\alpha^{-1/2} \, , \quad \alpha=1,\dots, N_i
\label{eqpp6}
\end{equation}
The nondimesional case is recovered for $E=m=1$ with ${\bf g}_\alpha=\boldsymbol{\gamma}_\alpha$,
$\alpha=1,\dots,N_i$.
Finally, let us consider the most general fluctuational
pattern, namely ${\mathbb D} {\mathbb I}_{m,E}^{(n,N_d,N_i)}(\{ \boldsymbol{\xi}_h^{(d)}(t) \}_{h=1}^{N_d}, 
\{ \boldsymbol{\xi}_\alpha^{(i)}(t) \}_{\alpha=1}^{N_i})$ accounting both for 
dissipative and fluid-inertial memory hydromechanics (${\mathbb D} {\mathbb I}$
is indeed for Dissipative-Inertial).
The ${\mathbb D} {\mathbb I}_{m,E}^{(n,N_d,N_i)}$ pattern can be obtained by merging the
dissipative pattern ${\mathbb D}_{m,E}^{(n,N_d,0)}$ with the inertial pattern ${\mathbb I}_{m,E}^{(n,\,N_i)}$,
setting to zero the instantaneous Stokes  friction. The equation of this pattern are not repeated as
they can be obtained  by summing the forces defined in the two aforementioned pattern  in the dynamics 
of ${\bf v}(t)$, and by considering both the $\{\boldsymbol{\theta}_h(t)\}_{h=1}^{N_d}$
and the $\{{\bf z}_\alpha(t)\}_{\alpha=1}^{N_i}$ degrees of freedom. Table \ref{tab1}
reports a synoptic reviews of the fluctuational patterns, and figure \ref{figblock} the
block-diagrammatic representation of the action  of the stochastic forcing and of the
auxiliary degrees of freedom in the dynamics of the different patterns.

The  pattern composition  can be  further formalized. Owing to the linearity in the
composition of forces, to the linearity of the fluid-dynamic interactions considered and to the
mutual independence of the  stochastic forcings acting on the different degrees of freedom,
an algebra of fluctuational patterns follows.
Let ${{{\mathbb D} {\mathbb I}^\prime}_{m,E}^{(n,N_d^\prime,N_i^\prime)}}$ and
 ${{{\mathbb D} {\mathbb I}^{\prime \prime}}_{m,E}^{(n,N_d^{\prime \prime},N_i^{\prime \prime})}}$.
two  fluctuational patterns associated with the same particle (of mass $m$) at the same temperature $T$
and corresponding to the action of different dissipative and fluid inertial kernels, 
$\{{\bf h}^\prime(t), {\bf k}^\prime(t)\}$ and $\{{\bf h}^{\prime \prime}(t), {\bf k}^{\prime \prime}(t) \}$ 
respectively defined by the hydromechanic systems of parameters
$\left ( \{ {\bf a}^\prime_h \}_{h=1}^{N_d^\prime}, \{ \lambda^\prime_h \}_{h=1}^{N_d^\prime}, 
\{ {\bf g}^\prime_\alpha \}_{\alpha=1}^{N_i^\prime}, \{ \mu^\prime_\alpha \}_{\alpha=1}^{N_i^\prime} 
\right )$ and
$\left ( \{ {\bf a}^{\prime \prime}_h \}_{h=1}^{N_d^{\prime \prime}}, \{ \lambda^{\prime \prime}_h \}_{h=1}^{N_d^{\prime \prime}},  
\{ {\bf g}^{\prime \prime}_\alpha \}_{\alpha=1}^{N_i^{\prime \prime}}, \{ \mu^{\prime \prime}_\alpha \}_{\alpha=1}^{N_i^{\prime \prime}} 
\right )$.

Let $\{ {\bf a}^\prime_h \}_{h=1}^{N_d^\prime}$, and $\{ {\bf a}^{\prime \prime}_k \}_{k=1}^{N_d^{\prime \prime}}$ two
vector sets. The {\em extension} $\{ {\bf a}_h \}_{h=1}^{N_d^\prime+N_d^{\prime \prime}}$ of the two sets, indicated by
\begin{equation}
\{ {\bf a}_h \}_{h=1}^{N_d^\prime+N_d^{\prime \prime}} = \{ {\bf a}^\prime_h \}_{h=1}^{N_d^\prime} \, \odot \{ {\bf a}^{\prime \prime}_k \}_{k=1}^{N_d^{\prime \prime}}
\label{eqpp7}
\end{equation}
is the vector set formed by the ordered  sequence of the elements of the two vector sets, i.e.
\begin{equation}
\{ {\bf a}_h \}_{h=1}^{N_d^\prime+N_d^{\prime \prime}}= \{ {\bf a}_1^\prime,\dots,{\bf a}_{N_d^\prime}^\prime,
{\bf a}_1^{\prime \prime} ,\dots,{\bf a}_{N_d^{\prime \prime}}^{\prime \prime} \}
\label{eqpp8}
\end{equation}
The following composition rule, referred to as pattern summation and indicated by ``$\oplus$'', can be
introduced amongst patterns,
\begin{eqnarray}
{{{\mathbb D} {\mathbb I}^\prime}_{m,E}^{(n,N_d^\prime,N_i^\prime)}}(
\{ \boldsymbol{\xi}_h^{\prime (d)} \}_{h=1}^{N_d^\prime},
\{ \boldsymbol{\xi}_\alpha^{\prime (i)} \}_{\alpha=1}^{N_i^\prime}) \oplus
 {{{\mathbb D} {\mathbb I}^{\prime \prime}}_{m,E}^{(n,N_d^{\prime \prime},N_i^{\prime \prime})}}(
\{ \boldsymbol{\xi}_h^{\prime \prime (d)} \}_{h=1}^{N_d^{\prime\prime}},
\{ \boldsymbol{\xi}_\alpha^{\prime \prime (i)} \}_{\alpha=1}^{N_i^{\prime \prime}}) \nonumber \\
= {\mathbb D} {\mathbb I}_{m,E}^{(n,N_d^\prime+N_d^{\prime \prime},N_i^\prime+N_i^{\prime \prime})}(  \{ \boldsymbol{\xi}_h^{\prime (d)} \}_{h=1}^{N_d^\prime}
\odot \{ \boldsymbol{\xi}_h^{\prime \prime (d)} \}_{h=1}^{N_d^{\prime\prime}},
\{ \boldsymbol{\xi}_\alpha^{\prime (i)} \}_{\alpha=1}^{N_i^\prime})
\odot
\{ \boldsymbol{\xi}_\alpha^{\prime \prime (i)} \}_{\alpha=1}^{N_i^{\prime \prime}})
\label{eqpp9}
\end{eqnarray}
where the pattern ${\mathbb D} {\mathbb I}_{m,E}^{(n,N_d^\prime+N_d^{\prime \prime},N_i^\prime+N_i^{\prime \prime})}$
is defined by the extension of all the vector sets of hydromechanic, fluctuational parameters, auxiliary variables
and stochastic forcings.
The  pattern $ {\mathbb D} {\mathbb I}_{m,E}^{(n,N_d^\prime+N_d^{\prime \prime},N_i^\prime+N_i^{\prime \prime})}$ so
constructed corresponds to the stochastic dynamics, consistent with FD1k, i.e. with the Langevin
condition for a particle  moving in a fluid, the hydromechanic response of which is associated
with the dissipative and fluid-inertial kernels ${\bf h}(t)$, ${\bf k}(t)$,
\begin{equation}
{\bf h}(t)= {\bf h}^\prime(t)+{\bf h}^{\prime \prime}(t) \, , \qquad {\bf k}(t)= {\bf k}^\prime(t)+{\bf k}^{\prime \prime}(t)  
\label{eqpp10}
\end{equation}
that is the superposition of the hydromechanic forces  characterizing
${{{\mathbb D} {\mathbb I}^\prime}_{m,E}^{(n,N_d^\prime,N_i^\prime)}}$ and 
${{{\mathbb D} {\mathbb I}^{\prime \prime}}_{m,E}^{(n,N_d^{\prime \prime},N_i^{\prime \prime})}}$.
The above composition algebra can be applied to generic patterns, so that,
${\mathbb S}_{m,E}^{(n,1,0)} \oplus {\mathbb D}_{m,E}^{(n,N_d,0)}$ or ${\mathbb D}_{m,E}^{(n,N_d,0)} \oplus
{\mathbb I}_{m,E}^{(n,1,N_i)}$ are elements of this algebra corresponding to specific fluid-particle interactions.
Observe that ${\mathbb D}_{m,E}^{(n,N_d,0)} \oplus
{\mathbb I}_{m,E}^{(n,1,N_i)}$ is different from ${\mathbb D}{\mathbb I}_{m,E}^{(n,N_d,N_i)}$ due to the
presence of an additional Stokesian   instantaneous dissipation term. Consequently, the latter can be
recovered as  ${\mathbb D}_{m,E}^{(n,N_d,0)} \oplus
{\mathbb I}_{m,E}^{(n,1,N_i)}|_{\boldsymbol{\eta}={\bf 0}}={\mathbb D} {\mathbb I}_{m,E}^{(n,N_d+1,N_i)}$, 
and the dissipative degrees of freedom
associated with the instantaneous dissipation characterizing ${\mathbb I}_{m,E}^{(n,1,N_i)}$ can be neglected,
so that $N_d+1$ in the definition of ${\mathbb D} {\mathbb I}_{m,E}^{(n,N_d+1,N_i)}$ reduces to $N_d$.
\subsection{Comment on the uniqueness of the representation}
\label{subsecx}
A final comment involves the uniqueness of the FD3k modal representation.
To address this issue, consider the case of purely dissipative memory,
adopting the nondimensional formulation used in Sections \ref{sec4}-\ref{sec6}
in the case  
the kernel $h(t)$ is  expressed in the form
\begin{equation}
h(t)= \sum_{h=1}^{N_d} \sum_{k=1}^{N_d} 
\alpha_h \left ( e^{-{\boldsymbol \Lambda} \, t} \right )_{hk}
 \beta_k
\label{eq11_2}
\end{equation}
where ${\boldsymbol \Lambda}$ is a $N_d \times N_d$ positive definite matrix,
and $\alpha_h$, $\beta_k$ are such that $h(t) >0$, for any $t>0$.
Of course, the kernel $h(t)$ expressed by eq. (\ref{eq11_2}) can be also
expressed in the form eq. (\ref{eq3_1}), where $\lambda_i$, $i=1,\dots, N_d$ are the
eigenvalues of ${\boldsymbol \Lambda}$, but the aim of the present analysis is to
enforce the representation eq. (\ref{eq11_2})  in the derivation of the FD3k modal
representation.
Eq. (\ref{eq11_2}) can be  compactly rewritten   as
\begin{equation}
h(t)= \left ( {\boldsymbol \alpha}, e^{-{\boldsymbol \Lambda} \, t} \, {\boldsymbol \beta} \right )
\label{eq11_3}
\end{equation}
where $({\boldsymbol \alpha},{\boldsymbol \beta})=\sum_{h=1}^{N_d} \alpha_h \, \beta_h$.
Within this setting, it is natural
to express the internal dissipative degrees of freedom ${\boldsymbol \theta}(t)$
in the form
\begin{equation}
{\boldsymbol \theta}(t)= e^{-{\boldsymbol \Lambda} \, t} * {\boldsymbol \beta} \, v(t)
\label{eq11_4}
\end{equation}
or, componentwise
\begin{equation}
\theta_h(t)= \sum_{k=1}^{N_d} \left ( e^{-{\boldsymbol \Lambda} \, t} \right )_{h,k} \, \beta_k \, v(t)
\label{eq11_5}
\end{equation}
leading to the  dynamic representation of the FD3k relation 
\begin{eqnarray}
\frac{d v}{d t} & = & - \sum_{h=1}^{N_d} \alpha_h \, \theta_h \label{eq11_6} \\
\frac{d \theta_h}{d t} & = & - \sum_{k=1}^{N_d} \Lambda_{hk} \, \theta_k + \beta_h \, v
+ \sqrt{2} \, \sum_{k=1}^{N_d} \sigma_{hk} \, \xi_k(t)
\nonumber
\end{eqnarray}
where $\sigma_{hk}$ are the entries of a $N_d \times N_d$ matrix
${\boldsymbol \sigma}$ to be determined.
The associated Fokker-Planck equation reads
\begin{eqnarray}
\frac{\partial p}{\partial t} & = & \sum_{h=1}^{N_d} \frac{\partial p}{\partial v} +
\sum_{h=1}^{N_d} \frac{\partial }{\partial \theta_h} \left ( \sum_{k=1}^{N_d} \Lambda_{hk} \, \theta_k \, p \right ) \nonumber \\
& - & \sum_{h=1}^{N_d} \beta_h \, v \, \frac{\partial p}{\partial \theta_h} + \sum_{h,k=1}^{N_d} D_{hk} \, \frac{\partial^2 p}{\partial \theta_h \partial \theta_k}
\label{eq11_7}
\end{eqnarray}
where $D_{hk}= \sum_{j=1}^{N_d} \sigma_{hj} \, \sigma_{kj}$. 
From moment analysis at steady steate, enforcing the condition $m_{vv}^*=1$, $m_{\theta_h v}^*=0$,
one obtains
\begin{equation}
2 D_{ij}= \sum_{k=1}^{N_d}  \left ( \Lambda_{ik} \, m_{\theta_j \theta_k}^* + 
\Lambda_{jk} m_{\theta_i \theta_k}^* \right )
\label{eq11_8}
\end{equation}
and
\begin{equation}
\sum_{h=1}^{N_d} \alpha_h \, m_{\theta_i \theta_h}^* = \beta_i
\label{eq11_9}
\end{equation}
The latter equation is satisfied by
\begin{equation}
m_{\theta_i \theta_h}^*= \frac{\beta_i}{\alpha_i} \, \delta_{ih}
\label{eq11_9bis}
\end{equation}
Inserting this expression into eq. (\ref{eq11_8}) the expression for
$D_{ij}$ follows
\begin{equation}
D_{ij}= \frac{1}{2} \left ( \Lambda_{ij} \, \frac{\beta_j}{\alpha_j}+ \Lambda_{ji} \, \frac{\beta_i}{\alpha_i} \right )
\label{eq11_10}
\end{equation}
which is a symmetric and positive definite  matrix ${\bf D}$.  Consequently, there exists
a unique symmetric and positive definite matrix ${\boldsymbol \sigma}$, such that
${\boldsymbol \sigma}={\bf D}^{1/2}$.

The  principal conclusions that can be drawn from this example are the following:
(i)  the explicit representation of the modal FD3k  relation is not unique, 
as in the present case
we could have used the simpler  form derived in section \ref{sec4}, in which each mode
is characterized by an independent fluctuational forcing,
 nevertheless (ii) the set of all these equivalent FD3k representations possesses analogous
structural properties in the meaning of the block-diagrammatic structure depicted in figure
\ref{figblock} panel (b), and corresponding to
a well defined fluctuational pattern, in this case the ${\mathbb D}$-pattern,
where $\{ \xi_k(t)\}_ {k=1}^{N_d}$
interact with each dissipative degree of freedom $\theta_h(t)$,
$h=1,\dots,N_d$
Moreover, independently of
the fine modal representation of the fluctuational forcing, the dissipative internal degrees of
freedom are at steady state (equilibrium) uncorrelated  from each other, as it follows
from the diagonal structure of the moments $m_{\theta_i \theta_h}^*$ at equilibrium, eq. (\ref{eq11_9bis}).

To conclude, the fluctuational patterns considered above refer to
a diagonal representation of the internal degrees of freedom. 
Non-diagonal internal dynamics can be introduced without
problem, still representing a fully equivalent model, 
without modifying the structural properties of the fluctuational patterns.

\section{Fluctuation-dissipation theory in confined systems}
\label{sec9}

The extension of fluctuation-dissipation theory from the free-space to
confined systems, such as microchannels and, more generally, any fluidic 
device for which its characteristic size is comparable with that
of the diffusing particles, requires particular care and attention,
as it poses new challenging problems that are not encountered in the
free-space case.
The main hydromechanic effect induced by confinement is that the
fluid-particle interactions depend on the particle location
in the proximity of the device walls, and consequently they become
function of the particle position \cite{procopiofluids,brenner}.

This poses a conceptual  challenge in  expressing the particle equation
of motion, even  in the case the thermal fluctuations are overlooked,
whenever memory effects, be them associated with fluid-inertial interactions
or with the non-Newtonian rheology of the medium, are accounted for.
This stems from the intrinsic ambiguity in the functional
form of the force exerted by the fluid on the
particle  that can be derived from the hydrodynamic analysis
of the problem.  

For the sake of physical clearness, in this section we use
dimensional units and vector-valued quantities for the particle
velocity.

Indeed, in the  case of the Stokes regime, the fluctuation-dissipation
patterns derived in the free-space can be straingthforwardly
extended to the confined case. If $\boldsymbol{\eta}({\bf x})$ is
the friction tensor function of particle position ${\bf x}$,
the fluctuational pattern is 
${\mathbb S}_{m,E}^{(3,1,0)}(\boldsymbol{\xi}(t))$ with
${\bf a}({\bf x}) = \sqrt{ k_B \, T} \boldsymbol{\eta}^{1/2}({\bf x})$,
where the symmetric   square-root matrix $\boldsymbol{\eta}^{1/2}({\bf x})$
is uniquely defined  since $\boldsymbol{\eta}({\bf x})$ is symmetric and
positive definite \cite{procopiofluids,brenner}.
This corresponds to the equation of motion 
\begin{equation}
m \, \frac{d {\bf v}(t)}{d t}= - \boldsymbol{\eta}({\bf x}(t)) \, {\bf v}(t) +
\sqrt{2 \, k_B \, T} \, \boldsymbol{\eta}^{1/2}({\bf x}(t)) \, \boldsymbol{\xi}(t)
\label{eqco1}
\end{equation}
where $\boldsymbol{\xi}(t)$ is the distributional
derivative of a three-dimensional   Wiener process, and eq. (\ref{eqco1})
is coupled with the kinematic equation
\begin{equation}
\frac{d {\bf x}(t)}{d t}=  {\bf v}(t)
\label{eqco2}
\end{equation}

In order to understand the hydromechanic complexity of the extension of
this results in the presence of memory interactions,
consider the case of a rigid particle of mass $m$ moving with
velocity ${\bf v}(t)$ and position vector ${\bf x}(t)$
in a fluid device defined by the spatial domain $\Omega$, with a non-moving
solid boundary $S_w$. Let $S_p({\bf x}(t))$ be the external surface
of the particle, the spatial location of which depends on the
particle position ${\bf x}(t)$, and assume, without loss
of generality, no slip boundary conditions both on $S_w$ and $S_p$.
As regards fluid hydrodynamics and rheological properties, 
consider  the incompressible time-dependent Stokes equation, and
Newtonian constitutive equations (viscosity $\mu$, fluid density $\rho$).
In order to avoid confusion, since ${\bf x}(t)$ refers to the particle
position, the running spatial variables for the
hydrodynamic fields in $\Omega$ will be indicated with ${\bf y}$,
and correspondingly  the differential operators will refer to this variable.
Thus, the fluid velocity field  ${\bf u}({\bf y},t)$ is
the solution of the moving-boundary problem
\begin{eqnarray}
\rho \, \frac{\partial {\bf u}({\bf y},t)}{\partial t} & = &
\mu \, \nabla^2 {\bf u}({\bf y},t) - \nabla p({\bf y},t) = -
\nabla \cdot {\bf T}({\bf y},t) \, , \quad \nabla \cdot {\bf u}({\bf y},t) = 0
\, , \quad {\bf y} \in \Omega \nonumber \\
{\bf u}({\bf y},0)|_{{\bf y} \in \Omega} & = & 0 \, , \quad {\bf u}({\bf y},t) |_{{\bf  y} \in S_w}=0 \, ,
\quad {\bf u}({\bf y},t)|_{{\bf y} \in S_p({\bf x}(t))}= {\bf v}(t)
\label{eqco3}
\end{eqnarray}
where ${\bf T}({\bf y},t)$ is the overall stress tensor accounting both for
pressure ans shear stress contributions.
The hydromechanics of the problem is defined by the force exerted by the
fluid onto the particle
\begin{equation}
{\bf F}_{f \rightarrow p}[{\bf v}(t);{\bf x}(t)]= - \int_{S_p({\bf x}(t))}
{\bf T}({\bf y},t) \cdot {\bf n}_e({\bf y},t) \, d S({\bf y})
\label{eqco4}
\end{equation}
where ${\bf n}_e({\bf y},t)$ is the normal, outwardly oriented, unit vector
to $S_p({\bf x}(t))$ at the point ${\bf y} \in S_p({\bf x}(t))$.
Observe that eq. (\ref{eqco3}) defines a moving-boundary problem due to
explicit dependence of the boundary conditions on the particle position
within $\Omega$, and because of it, also the force defined by eq. (\ref{eqco4})
depends on ${\bf x}(t)$.

While  this moving boundary problem can be suitably approached via
numerical simulations, the analytic determination of the
mean-field hydromechanic force eq. (\ref{eqco4}) shows  intrinsic
criticalities. Owing to the linearity of eqs. (\ref{eqco3})-(\ref{eqco4}),
the hydrodynamic problem can be solved in the Laplace domain at any fixed
value of ${\bf x}$, i.e. taking the particle
position as a parameter. Letting $\widehat{\bf v}(s)=L[{\bf v}(t)]=
\int_0^\infty e^{-s \, t}  \, {\bf v}(t) \, d t$ be the Laplace transform
of the particle velocity, and  
\begin{equation}
\widehat{\bf F}_{f \rightarrow p}[{\widehat {\bf v}}(s);{\bf x}]=
- \int_{S_p({\bf x})}
\widehat{\bf T}({\bf y},s) \cdot {\bf n}_e({\bf y}) \, d S({\bf y})
\label{eqco5}
\end{equation}
the Laplace transform of the force on the particle at the fixed
location ${\bf x}$, the latter quantity is amenable to
analytical  calculation, obtaining  by linearity
\begin{equation}
\widehat{\bf F}_{f \rightarrow p}[\widehat{\bf v}(s);{\bf x}]=
- \widehat{\bf K}(s;{\bf x}) \, \widehat{\bf v}(s)
\label{eqco6}
\end{equation}
But, even if a closed and exact expression for $\widehat{\bf F}_{f \rightarrow p}[\widehat {\bf v}(s);{\bf x}]$, or for  its
inverse Laplace transform in the time domain would be available,
and a variety of semianalytical models are available in the literature \cite{std1,std2,std3}, this
analytical result
would not solve the hydromechanic problem as regards  the
determination of ${\bf F}_{f \rightarrow p}[{\bf v}(t);{\bf x}(t)]$ in
eq. (\ref{eqco4}), just because the latter is a functional of the particle positional
history $\{{\bf x}(\tau)\}_{\tau=0}^t$ from $\tau=0$  to $\tau=t$, 
evolving according to eq.
(\ref{eqco2}) while in eq. 
(\ref{eqco6}) ${\bf x}$ is a fixed parameter.

This problem has been analyzed in \cite{gppfd} in the case of purely
dissipative viscoelastic kernels, i.e. neglecting fluid-inertial 
contributions. In this case, for fixed ${\bf x}$, we have
\begin{equation} 
\widehat{\bf F}_{f \rightarrow p}[{\widehat {\bf v}}(s);{\bf x}]
= - \widehat{\bf h}(s;{\bf x}) \, \widehat{\bf v}(s) \quad
\Rightarrow   {\bf F}_{f \rightarrow p}[{\bf v}(t);{\bf x}]
= - \int_0^t {\bf h}(t-\tau;{\bf x}) \, {\bf v}(\tau) \, d \tau 
\label{eqco7}
\end{equation}
It has been shown in \cite{gppfd} that fluctuation-dissipation
theory, i.e. the analysis of the hydromechanic problem
in the presence of thermal fluctuations, provides a way
for obtaining thermodynamically
consistent expression for the functional form of the
mean-field force acting onto the particle
${\bf F}_{f \rightarrow p}[{\bf v}(t);{\bf x}(t)]$, and the
explicit representation of the stochastic thermal force ${\bf R}(t)$.

The fluctuation-dissipation theory developed in \cite{gppfd}
enforces two basic representation principles: (i)
the principle of  local realizability, corresponding to
the series expansion in exponentially decaying modes,
introduced in \cite{gpp_localrepr} and discussed in the introduction and in 
Section \ref{sec3}, and (ii) the principle
of local consistency.
Local realizability implies that the kernel ${\bf h}(t;{\bf x})$
for fixed ${\bf x}$, can be represented as
\begin{equation}
{\bf h}(t;{\bf x}) = \sum_{h=1}^{N_d} {\bf A}_h({\bf x}) \, \lambda_h \, e^{-\lambda_h \, t}
\label{eqco8}
\end{equation}
where ${\bf A}_h({\bf x})$ are symmetric, positive definite matrices
function of ${\bf x}$, while the relaxation rates are assumed to be position independent.
The principle of local consistency implies  that the
correct functional form for  the
force functional ${\bf F}_{f \rightarrow p}[{\bf v}(t);{\bf x}(t)]$ if
restricted to the case where ${\bf x}(t)={\bf x}^*=\mbox{const.}$,
should  coincide with ${\bf F}_{f \rightarrow p}[{\bf v}(t);{\bf x}^*]$
(in the present case defined by eq. (\ref{eqco7})), corresponding to
a fluctuation-dissipation problem analogous to those defined
in the free-space, for which the expression for the associated
effective thermal force ${\bf R}(t,{\bf x}^*)$ can be derived
using the fluctuational patterns addressed above (see also Appendix \ref{app1}).
Essentially, the local consistency principle,  restricts the
class of admissible expressions for ${\bf F}_{f \rightarrow p}[{\bf v}(t);{\bf x}(t)]$ but does not solve the problem.

The  solution of the problem can be obtained by enforcing    thermodynamic
principles,  namely the  property that the marginal density function 
for particle position in a closed and bounded domain $\Omega$, and in the
absence of
external forces and perturbations, should be uniform at equilibrium.
For the sake of exactness, this marginal density should be uniform in the
subset $\Omega^\prime \subseteq \Omega$ defined by accounting for the
possible sterical constraints associated with the finite size of the particles.
A stronger form of the
equilibrium principle states that  the uniformity condition
at equilibrium holds for  all the auxiliary variables
introduced in the local representation of the 
 memory dynamics.

The following result has been obtained in \cite{gppfd}. \\

\noindent
{\bf Theorem I } -  For the confined hydromechanic problem  at
constant temperature $T$ defined
by the position dependent memory kernel ${\bf h}(t;{\bf x})$ at constant
${\bf x}$ eq. (\ref{eqco8}), 
the only thermodynamically consistent representation of the
particle dynamics,  providing uniform equilibrium marginal position
density in a closed and bounded domain, and uniform statistical
properties for the auxiliary memory varables is given
by
\begin{eqnarray}
\frac{d {\bf x}(t)}{d t} & = & {\bf v}(t) \nonumber \\
m \, \frac{d {\bf v}(t)}{ d t} & = & - \sum_{h=1}^{N_d} {\bf A}_h^{1/2}({\bf x}(t)) \, \lambda_h \, {\bf z}_h
\label{eqco9} \\
\frac{d {\bf z}_h(t)}{d t} & = & -\lambda_h \, {\bf z}_h(t) + {\bf A}_h^{1/2}({\bf x}(
t)) \, {\bf v}(t) + \sqrt{2 \, k_B \, T} \, \boldsymbol{\xi}_h(t) 
\, , \qquad h=1,\dots,N_d\nonumber
\end{eqnarray}
where $\boldsymbol{\xi}_h(t)=(\xi_{h,1}(t),\xi_{h,2}(t),\xi_{h,3}(t))$ are 
the distributional
derivatives of  three-dimensional independent Wiener processes,
$\langle \xi_{h,i}(t) \, \xi_{k,j}(t^\prime) \rangle =
\delta_{hk} \, \delta_{ij} \, \delta(t-t^\prime)$, $h,k=1,\dots,N_d$, $i,j=1,2,3$.    \\

\noindent
For the proof see \cite{gppfd}. $\diamond$ \\

In the remainder  we extend this result to fluid inertial effects first
in a Newtonian fluid, and subsequently in the
 case of generic complex fluids characterized by a viscoelastic memory
response.

For a confined Newtonian fluid in the time-dependent incompressible Stokes
regime, the particle hydromechanics at constant particle position ${\bf x}$ is
defined by the equation
\begin{equation}
m \, \frac{d {\bf v}(t)}{d t}  =  -   \boldsymbol{\eta}({\bf x}) \, {\bf v}(t) -
{\bf k}(t;{\bf x}) * \left ( \frac{d {\bf v}(t)}{d t} + {\bf v}(0) \, \delta(t)
\right ) + {\bf R}(t;{\bf x})
\label{eqco10}
\end{equation}
where  $\boldsymbol{\eta}({\bf x})$ and  ${\bf k}(t;{\bf x})$ are respectively the friction tensor and the fluid-inertial tensorial
kernels  at constant ${\bf x}$. 
In eq. (\ref{eqco10}) we have neglected the added mass. This
contribution
can be easily imbedded in the theory as developed in  Appendix  \ref{app3}.
As in section \ref{sec2}, we assume 
 ${\bf k}(0,{\bf x})$ to be bounded, as it follows from the finite propagation
velocity of the shear stresses. 
 As for eqs. (\ref{eq2_8}) and (\ref{eq6_3}),  eq. (\ref{eqco10}) can be
transformed into
\begin{equation}
m \, \frac{d {\bf v}(t)}{ dt}= - \boldsymbol{\eta}({\bf x}) \, {\bf v}(t)
- {\bf k}(0;{\bf x}) \, {\bf v}(t) - \int_0^t \frac { d {\bf k}(t-\tau;{\bf x})}{d \tau} \, {\bf v}(\tau) \, d \tau + {\bf R}(t;{\bf x})
\label{eqco11}
\end{equation}
From the principle of local realizability (i.e. using  
the expansions outlined in section \ref {sec3}), we can represent
the kernel ${\bf k}(t;{\bf x})$ in the form
\begin{equation}
{\bf k}(t;{\bf x}) = \sum_{\alpha=1}^{N_i} {\bf g}_\alpha({\bf x}) \, e^{-\mu_\alpha \, t}
\label{eqco12}
\end{equation}
where ${\bf g}_\alpha({\bf x})$ are $3 \times 3$ matrices parametrically
dependent on ${\bf x}$, and the relaxation rates $\mu_\alpha$  are constant.
Substituting eq. (\ref{eqco12}) into eq. (\ref{eqco11}) we  obtain
\begin{equation}
m \, \frac{d {\bf v}(t)}{d t}= - \boldsymbol{\eta}({\bf x}) \, {\bf v}(t)
- {\bf G}({\bf x}) {\bf v}(t) + \sum_{\alpha=1}^{N_i} {\bf g}_\alpha({\bf x}) \,
\mu_\alpha \, e^{-\mu_\alpha t} * {\bf v}(t) + {\bf R}(t;{\bf x})
\label{eqco13}
\end{equation}
where ${\bf G}({\bf x})= \sum_ {\alpha=1}^{N_i} {\bf g}_\alpha({\bf x})$.
Eq. (\ref{eqco13}) admits a local representation, introducing the
auxiliary variables ${\bf z}_\alpha(t)=(z_{\alpha,1}(t),z_{\alpha,2}(t),z_{\alpha,3}(t))$, $\alpha=1,2,3$,
\begin{eqnarray}
m \, \frac{d {\bf v}(t)}{d t} & = & - \boldsymbol{\eta}({\bf x}) \, {\bf v}(t) -
{\bf G}({\bf x}) \, {\bf v}(t) + \sum_{\alpha=1}^{N_i} \mu_\alpha \, 
\boldsymbol{\gamma}_\alpha({\bf x}) \, {\bf z}_\alpha(t) + \sqrt{2} \, {\bf a}({\bf x}) \, \boldsymbol{\xi}(t) + \sqrt{2} \sum_{\alpha=1}^{N_i} {\bf d}_\alpha({\bf x}) \, \boldsymbol{\xi}_\alpha^\prime(t)
\nonumber \\
\frac{d {\bf z}_\alpha(t)}{d t} & = & - \mu_\alpha \, {\bf z}_\alpha(t)+ \boldsymbol{\zeta}_\alpha({\bf x}) \, {\bf v}(t) + \sqrt{2} \, {\bf c}_\alpha({\bf x}) \, 
\boldsymbol{\xi}_\alpha^\prime(t) \, , \quad \alpha=1,\dots,N_i
\label{eqco14}
\end{eqnarray}
where $\langle \xi_i(t) \, \xi_{\alpha,j}^\prime(t^\prime) \rangle=0$,
$\langle \xi_{i,\alpha}^\prime(t) \, \xi_{\beta,j}^\prime(t^\prime) \rangle
= \delta_{\alpha\beta} \, \delta_{ij} \, \delta(t-t^\prime)$, and
\begin{equation}
\boldsymbol{\gamma}_\alpha({\bf x}) \, \boldsymbol{\zeta}_\alpha({\bf x})=
{\bf g}_\alpha({\bf x}) \, , \qquad \alpha=1,\dots,N_i
\label{eqco15}
\end{equation}
We have he following results:\\

\noindent
{\bf Theorem II} - The equilibrium solution of the particle
dynamic equations
\begin{eqnarray}
\frac{d {\bf x}(t)}{d t} & = & {\bf v}(t) \nonumber \\
m \, \frac{d {\bf v}(t)}{d t} & = & - \boldsymbol{\eta}({\bf x}(t)) -
{\bf G}({\bf x}(t)) \, {\bf v}(t) + \sum_{\alpha=1}^{N_i} \mu_\alpha \,
\boldsymbol{\gamma}_\alpha({\bf x}(t)) \, {\bf z}_\alpha(t) + \sqrt{2} \, {\bf a}({\bf x}(t)) \, \boldsymbol{\xi}(t) + \sqrt{2} \sum_{\alpha=1}^{N_i} {\bf d}_\alpha({\bf x}(t)) \, \boldsymbol{\xi}_\alpha^\prime(t)
\nonumber \\
\frac{d {\bf z}_\alpha(t)}{d t} & = & - \mu_\alpha \, {\bf z}_\alpha(t)+ \boldsymbol{\zeta}_\alpha({\bf x}(t)) \, {\bf v}(t) + \sqrt{2} \, {\bf c}_\alpha({\bf x}(t)) \,
\boldsymbol{\xi}^\prime_\alpha(t) \, , \quad \alpha=1,\dots,N_i
\label{eqco16}
\end{eqnarray}
in a closed and bounded domain admits a uniform spatial marginal
distribution if 
\begin{equation}
\boldsymbol{\gamma}_\alpha({\bf x}) = \boldsymbol{\zeta}_\alpha({\bf x})=
{\bf g}_\alpha^{1/2}({\bf x}) \, , \qquad \alpha=1,\dots,N_i,
\label{eqco17}
\end{equation}
and
\begin{equation}
{\bf d}_\alpha({\bf x}) = - \sqrt{k_B \, T} \, {\bf g}_\alpha^{1/2}({\bf x}) \,,
\quad
{\bf c}_\alpha({\bf x}) = \sqrt{k_B \, T} \, {\bf I}
\quad \alpha=1,\dots,N_i
\label{eqco18}
\end{equation}
where ${\bf I}$ is the identity matrix.
Moreover, the statistical properties of the auxiliary ${\bf z}_\alpha$-variables
are uniform at equilibrium if and only if eqs. (\ref{eqco17}) are satisfied.
$\diamond$.\\

\noindent
{\bf Theorem III} - In the thermodynamically consistent case, i.e.
if eq. (\ref{eqco17}) are fulfilled, the force exerted by the
fluid onto the particle and the thermal force attain the
functional representation
\begin{equation}
{\bf F}_{f \rightarrow p}[{\bf v}(t),{\bf x}(t)]=
- \boldsymbol{\eta}({\bf x}(t)) \, {\bf v}(t) - {\bf G} ({\bf x}(t)) \, {\bf v}(t) + \sum_{\alpha=1}^{N_i} {\bf g}_\alpha^{1/2}({\bf x}(t) \, \mu_\alpha 
\int_0^t e^{-\mu_\alpha \, (t-\tau)} {\bf g}_\alpha^{1/2}({\bf x}(\tau)) \, {\bf v}(\tau)
\, d \tau 
\label{eqco19}
\end{equation}
\begin{equation}
{\bf R}(t;{\bf x}(t))= \sqrt{2 \, k_B \, T} \left [
 \boldsymbol{\eta}^{1/2}({\bf x}(t)) \, \boldsymbol{\xi}(t) - \sum_{\alpha=1}^{N_i} {\bf g}_\alpha^{1/2}({\bf x}(t)) \, \xi_\alpha^\prime(t) + \int_0^t
e^{-\mu_\alpha \, (t-\tau)} \, \xi_\alpha^\prime(\tau) \, d \tau \right ]
\label{eqco20}
\end{equation}
For the proofs of Thorems II and III see Appendix \ref{app3}. $\diamond$. \\

\vspace{0.2cm}

The results formalized by Theorems II and III in the presence of  fluid-inertial
 effects
are qualitatively and functionally analogous to those obtained in the dissipative case. The contribution of each ${\bf g}_\alpha({\bf x})$
is splitted symmetrically  onto the dynamic equation
for particle velocity and for the auxiliary memory variables
${\bf z}_\alpha(t)$,  conditions eq. (\ref{eqco16}).
Moreover, the resulting stochastic force eq. (\ref{eqco20})
depends on the actual particle position, i.e. on ${\bf x}(t)$ at time $t$
and not on its previous history.

Also in the confined case the concept of fluctuational patterns and
their algebra applies. Different fluctuational patterns can be combined together
for producing consistent models of particle hydrodynamics of increasing
hydrodynamic complexity owing to the independence of the stochastic
forcings acting on each internal degree of freedom.
Therefore, joining together the results of Theorems I, II and III,  the
general case of FD3k for confined particle hydromechanics in the presence of
dissipative and fluid inertial memory effects can be  derived.

\section{Concluding remarks}
\label{sec10}

A complete formulation of stochastic hydromechanics for generic
linear fluid and flow conditions has been developed, including 
the case of constrained geometries.
The approach followed starts from the solution of the FD3k relation,
i.e. the representation of the thermal force $R(t)$  in terms
of elementary stochastic processes, and derives from it
FD2k, i.e. its autocorrelation properties.  Eqs. (\ref{eq9_7})
and (\ref{eq9_7a}) solve  completely this problem
in the presence of fluid inertial effects.
It is remarkable   the occurrence of the impulsive contribution
$k(0) \, \delta (t)$ in these relations, 
that  mathematically implies the condition of bounded
$k(0)$.
This is exactly the result derived in \cite{gionavisco} for generic fluids
enforcing the boundedness of the velocity of propagation of the internal shear stresses.

It is  also worth of attention, the new, active role of fluctuation-dissipation
theory and thermodynamic concepts in deriving in the presence of
constraints and memory hydromechanics, a thermodynamically
consistent expression for the mean-field force alongside with
the expression for the thermal one. It suggests the importance of considering
hydromechanics in a fluctuational environment as the most conclusive
setting for deriving pure hydromechanic consequences.

The decomposition in internal degrees of freedom, possessing the property of being uncorrelated
with each other at equilibrium, provides an elegant 
way of framing FD3k in a compact way via the concept of
fluctuational patterns and their algebra. This concept is
of practical application to any problem of Brownian
motion theory and can be extended to microfluidic investigation including
the effects of external potentials and external velocity fields.
The  properties of the fluctuational patterns associated
with different hydrodynamic regimes is extensively used in part III
\cite{part2} to analyze the local regularity properties
of velocity fluctuations in a hydrodynamic perspective.
Morover, the structure of the fluctuational patterns  provides
a proper setting for generalizing the Kubo fluctuation-dissipation theory.

This point deserves some clarifications. Extensive hydrodynamic analysis,  coupled
with recent experimental results on the
possible violation of equipartition theorem in Brownian motion in liquids,
\cite{optics_express} suggests a generalization of the current
fluctuation-dissipation approach removing the Langevin
condition eq. (\ref{eq1_2}). Enforcing the structure
of the fluctuational patterns this can be naturally achieved
preserving the global fluctuation-dissipation relations.
This theoretical extension is thoroughly addressed in part III \cite{part3}.

Throughout this work we have considered the representation
 of the modal fluctuational
forces  proportional to the distributional derivatives of independent 
Wiener processes,
$\{\xi_h \}_{h=1}^{N_d}$, $\{ \xi_\alpha^\prime(t) \}_{\alpha=1}^{N_d}$. Of course, this
corresponds to the most traditional choice which, as discussed in \cite{part2},  leads  necessarily to
Gaussian equilibrium density function. Other choices  
are possible, still preserving the validity of FD1k, i.e. of eq. (\ref{eq1_3}).
This issue and its statistical physical implications are addressed in  part II
\cite{part2}.\\

\vspace{0.2cm}

\noindent
{\bf Acknowledgment - } This research  received financial support from ICSC---Centro Nazionale di Ricerca in High Performance Computing, Big Data and Quantum Computing, funded by European Union---NextGenerationEU. One of the authors (M.G.) is grateful to
M. G.  Raizen for precious discussions.

\appendix
\section{Vectorial models}
\label{app1}
In this section we consider the representation of fluctuation-dissipation
relations for vectorial models in which the hydrodynamic parameters are 
expressed
by non-isotropic matrices. A non-dimensional formulation is
adopted, where  ${\bf v}=(v_1,v_2,v_3)$ is the nondimensional
particle velocity fulfilling the
normalized equilibrium relation  $\langle v_i v_j \rangle_{\rm eq}= \delta_{ij}$. 

\subsection{Dissipative memory}
\label{sec8_1}
To begin with, consider a single dissipative 
mode with rate constant $\lambda$
\begin{equation}
\frac{d {\bf v}(t)}{d t}= -{\bf a} \, e^{-\lambda \, t} * {\bf v}(t)
+ {\bf R}(t)
\label{eq8_1}
\end{equation}
where ${\bf v}(t)=(v_1(t),v_2(t),v_3(t))$, and similarly for ${\bf R}(t)$, while
${\bf a}=(a_{ij})_{i,j=1}^3$ is the nondimensional friction matrix,
which is symmetric and positive definite. Introducing $\boldsymbol{\theta}(t)=
e^{-\lambda \, t} * {\bf v}(t)$, and assuming, as in the main text, 
that the stochastic
fluctuations act solely on the internal degrees of freedom, we have
componentwise,
\begin{eqnarray}
\frac{d v_i}{d t} & = & - \sum_{j=1}^3 a_{i,j} \, \theta_j  \nonumber \\
\frac{d \theta_i}{d t} & =  & - \lambda \, \theta_i + v_i + \sqrt{2} \, \sum_{j=1}^3
\sigma_{i,j} \, \xi_j(t) \label{eq8_2}
\end{eqnarray}
$i=1,2,3$, where $\sigma_{i,j}$ are the entries of the 
symmetric matrix $\boldsymbol{\sigma}=\boldsymbol{\sigma}^T$ to be determined, and  $\langle \xi_i(t) \, \xi_j(t^\prime) \rangle =
\delta_{ij} \, \delta(t-t^\prime)$. The Fokker-Planck equation for the
density $p({\bf v}, \boldsymbol{\theta},t)$ associated with eq. (\ref{eq8_2})
reads
\begin{eqnarray}
\frac{\partial p}{\partial t}= \sum_{i=1}^3 \frac{\partial }{\partial v_i}
\left ( \sum_{j=1}^3 a_{ij} \, \theta_j \, p \right )
+ \sum_{i=1}^3 \frac{\partial }{\partial \theta_i}
\left ( \lambda \, \theta_i \, p \right ) -
\sum_{i=1}^3 v_i  \frac{\partial p}{\partial \theta_i} + \sum_{i,j=1}^3
D_{ij} \, \frac{\partial^2 p}{\partial \theta_i \partial \theta_j}
\label{eq8_3}
\end{eqnarray}
where
\begin{equation}
D_{ij}= \sum_{h=1}^3 \sigma_{ih} \, \sigma_{jh}
\label{eq8_4}
\end{equation}
The conditions to be imposed to the moments at steady-state are
\begin{equation}
m_{v_i v_j}^*= \delta_{ij} \, , \qquad m_{\theta_i v_j}^*=0 \, ,
\qquad i,j=1,2,3
\label{eq8_5}
\end{equation}
The dynamic equations for the moments  read
\begin{equation}
\frac{d m_{v_h v_k}}{d t} = - \sum_{j=1}^3 a_{hj}  \, m_{\theta_j v_k}
- \sum_{j=1}^3 a_{kj} \, m_{\theta_j v_h}
\label{eq8_6}
\end{equation}
that is identically verified at equilibrium, owing to eq. (\ref{eq8_5}).
Similarly,
\begin{equation}
\frac{d m_{\theta_h \theta_k}}{d t}= -2 \, \lambda \, m_{\theta_h \theta_k} - m_{\theta_h v_k} - m_{\theta_k v_h} + 2 \, D_{hk}
\label{eq8_7}
\end{equation}
At equilibrium, it reduces to
\begin{equation}
m_{\theta_h \theta_k}^*= \frac{D_{hk}}{\lambda}
\label{eq8_8}
\end{equation}
Finally, for the mixed moments $m_{\theta_k v_h}$ we have
\begin{equation}
\frac{d m_{\theta_k v_h}}{d t} = - \sum_{j=1}^3 a_{hj} m_{\theta_j \theta_k} - \lambda \, m_{\theta_k v_h} + m_{v_h v_k}
\label{eq8_9}
\end{equation}
that at steady-state becomes 
\begin{equation}
\sum_{j=1}^3 a_{hj} \, m_{\theta_j \theta_k}^*=  m_{v_h v_k}^*
\label{eq8_10}
\end{equation}
Using eqs. (\ref{eq8_5}) and  (\ref{eq8_8}), eq. (\ref{eq8_10})
can be compactly expressed in matrix form
\begin{equation}
{\bf a} \, {\bf D} =\lambda
\label{eq8_11}
\end{equation}
Since ${\bf D}=\boldsymbol{\sigma}^2$, and owing to the  symmetric
and positive definite nature of ${\bf a}$, there exists a unique
symmetric and positive semidefinite  matrix 
 $\boldsymbol{\sigma}$ such that \cite{matrix}
\begin{equation}
\boldsymbol{\sigma} = \sqrt{\lambda} \,{\bf a}^{-1/2}
\label{eq8_12}
\end{equation}
This result can be generalized  to a system of dissipative modes
\begin{equation}
\frac{d {\bf v}(t)}{d t} = - \sum_{h=1}^{N_d} {\bf a}^{(h)} \, e^{-\lambda_h \, t} * {\bf v}(t) + {\bf R}(t)
\label{eq8_13}
\end{equation}
Letting $\boldsymbol{\theta}^{(h)}(t)= e^{-\lambda_h \, t} * {\bf v}(t)$,
$h=1,\dots,N_d$,  we
have the representation
\begin{eqnarray}
\frac{d {\bf v}}{d t} & = & - \sum_{h=1}^{N_d} {\bf a}^{(h)} \, \boldsymbol{\theta}^{(h)} \nonumber \\
\frac{d \boldsymbol{\theta}^{(h)}}{d t} & = &- \lambda_h \,
\boldsymbol{\theta}^{(h)} + {\bf v} + \sqrt{2} \boldsymbol{\sigma}^{(h)} \, \boldsymbol{\xi}^{(h)}
\label{eq8_14}
\end{eqnarray}
where $\boldsymbol{\sigma}^{(h)}$ are symmetric $3\times 3$ matrices,
$\boldsymbol{\xi}^{(h)}(t)=(\xi_i^{(h)}(t))_{i=1,3}$, and 
\begin{equation}
\langle \xi^{(h)}_i(t) \, \xi^{(k)}_j(t^\prime) \rangle = \delta^{hk} \delta_{ij} \delta(t-t^\prime)
\label{eq8_14bis}
\end{equation}
As the dissipative internal degrees of freedom are uncorrelated, one
finally obtains for the matrices $\boldsymbol{\sigma}^{(h)}$ the
expression
\begin{equation}
\boldsymbol{\sigma}^{(h)}= \sqrt{\lambda_h} \, \left ( {\bf a}^{(h)} \right )^{-1/2} \, , \quad h=1,\dots,N_d
\label{eq8_15}
\end{equation}
generalizing  eq. (\ref{eq8_12}).

\subsection{Inertial and dissipative memory}
\label{sec8_2}

Next, consider the simultaneous occurrence of dissipative and inertial
memory effects, in the presence of a single mode for the two phenomena,
\begin{equation}
\frac{d {\bf v}(t)}{d t} = - {\bf a} \, e^{-\lambda \, t} * {\bf v}(t) - \boldsymbol{\gamma} \,
{\bf v}(t) + \mu \, \boldsymbol{\gamma} \,
e^{-\mu \, t} * {\bf v}(t) + {\bf R}(t)
\label{eq8_16}
\end{equation}
Introducing the internal degrees of freedom $\boldsymbol{\theta}(t)=e^{-\lambda \, t} *  {\bf v}(t)$, ${\bf z}(t)= e^{-\mu \, t} * {\bf v}(t)$, eq.
(\ref{eq8_16}) can be represented as
\begin{eqnarray}
\frac{d v_i}{d t} &= &- \sum_{j=1}^3 a_{ij} \, \theta_j - \sum_{j=1}^3 \gamma_{ij} \, v_j + \mu \, \sum_{j=1}^3 \gamma_{ij} \, z_j + \sum_{j=1}^3 d_{ij} \,
\xi^\prime_j(t) \nonumber \\
\frac{d \theta_i}{d t} &= & - \lambda \, \theta_i + v_i + \sum_{j=1}^3 b_{ij} \, \xi_j(t) \label{eq8_17} \\
\frac{d z_i}{d t} & = & - \mu \, z_i + v_i + \sum_{j=1}^3 c_{ij} \, \xi_j^\prime(t) \nonumber
\end{eqnarray}
$i=1,2,3$, where $b_{ij}$, $d_{ij}$ and $c_{ij}$ are the entries of the
the $3 \times 3$ symmetric matrices ${\bf b}$, ${\bf d}$, and ${\bf c}$, 
respectively.
The Fokker-Planck equation for the density $p({\bf v},\boldsymbol{\theta},{\bf z},t)$ associated with eq. (\ref{eq8_17}) is
\begin{eqnarray}
\frac{\partial p}{\partial t} & = &  \sum_{i=1}^3 \frac{\partial }{\partial v_i} \left (
\sum_{j=1}^3 a_{ij} \, \theta_j \, p \right ) + \sum_{i=1}^3 \frac{\partial }{\partial v_i} \left ( \sum_{j=1}^3 \gamma_{ij} \, v_j \, p \right ) - \mu \sum_{i=1}^3 \frac{\partial }{\partial v_i} \left ( \sum_{j=1}^3 \gamma_{ij} \, z_j \, p \right )
\nonumber \\
& +  & \sum_{i,j=1}^3 D_{ij}^{(1)} \frac{\partial^2 p}{\partial v_i \partial v_j}
 +  2 \, D_{ij}^{(c)} \, \frac{\partial^2 p }{\partial v_i \partial z_j}
+ \lambda \, \sum_{i=1}^3 \frac{\partial }{\partial \theta_i} \left ( \theta_i \, p \right )
- \sum_{i=1}^3  \, \frac{\partial p}{\partial \theta_i} + \sum_{i,j=1}^3 D_{ij} \, \frac{\partial^2 p}{\partial \theta_i \partial \theta_j} 
\nonumber \\
& + & \mu \, \sum_{i=1}^3 \frac{\partial }{\partial z_i} \left ( z_i \, p \right )
-\sum_{i=1}^3 v_i \, \frac{\partial p}{\partial z_i}
+ \sum_{i,j=1}^3 D_{ij}^{(2)} \, \frac{\partial^2 p}{\partial z_i \partial z_j} 
\label{eq8_18}
\end{eqnarray}
where
\begin{equation}
{\bf D}= {\bf b}^2 \, , \quad {\bf D}^{(1)}= {\bf d}^2 \, ,  \quad {\bf D}^{(2)}
= {\bf c}^2 \, , \quad {\bf D}^{(c)}= \frac{ {\bf c} \, {\bf d}+ {\bf d} \, {\bf c}}{2}
\label{eq8_19}
\end{equation}
and $D_{i,j}$ are the entries of the $3 \times 3$ matrix ${\bf D}$,
and analogously for ${\bf D}^{(u)}$, $u=1,2,c$.
As in the scalar case, the conditions to be imposed at steady state are
\begin{equation}
m_{v_i v_j}^*= \delta_{ij} \, , \quad m_{\theta_i v_j}^*= m_{z_i v_j}^*=0 \, ,
\qquad i,j=1,2,3
\label{eq8_20}
\end{equation}
Consider the second-order moment equations:
\begin{eqnarray}
\frac{d m_{v_h v_k}}{ dt}  & = & - \sum_{j=1}^3 a_{hj} \, m_{\theta_j v_k} - \sum_{j?1}^3 a_{kj} \, m_{\theta_j v_h} - \sum_{j=1}^3 \gamma_{hj} \, m_{v_k v_j} - \sum_{j=1}^3 \gamma_{kj} \, m_{v_h v_j} \nonumber \\
& + & \mu \, \sum_{j=1}^3 \gamma_{hj} m_{z_j v_k} + \mu \, \sum_{j=1}^3 \gamma_{kj} m_{z_j v_h} + 2 \, D_{hk}^{(1)}
\label{eq8_21}
\end{eqnarray}
At equilibrium,
\begin{equation}
\gamma_{hk}+\gamma_{kh} = 2 \, D_{hk}^{(1)}
\label{eq8_22}
\end{equation}
Since $\gamma_{hk}$ is symmetric, we have
\begin{equation}
D_{hk}^{(1)} = \gamma_{hk} \, \quad \Rightarrow \quad {\bf d}= \boldsymbol{\gamma}^{1/2}
\label{eq8_23}
\end{equation}
Next,
\begin{equation}
\frac{ d m_{\theta_h \theta_k}}{d t}= - 2 \, \lambda \, m_{\theta_h \theta_k}- m_{\theta_h v_k} - m_{\theta_k v_h} + 2 \, D_{hk}
\label{eq8_24}
\end{equation}
At steady-state, enforcing the  conditions eq. (\ref{eq8_20}),
we have
\begin{equation}
m_{\theta_h \theta_k}^* = \frac{D_{hk}}{\lambda}
\label{eq_24}
\end{equation}
As regards $m_{z_h z_k}$,
\begin{equation}
\frac{ d m_{z_h z_k}}{d t} = - 2 \, \mu \, m_{z_h z_k} - m_{x_h v_k} - m_{z_k v_h} + 2 \, D_{hk}^{(c)}
\label{eq8_25}
\end{equation}
that, at equilibrium, provides
\begin{equation}
m_{z_h z_k}^*= \frac{ D_{hk}^{(2)}}{\mu}
\label{eq8_26}
\end{equation}
The evolution equation of the mixed moments associated with
dissipative/inertial degrees of freedom is given by
\begin{equation}
\frac{d m_{\theta_h z_k}}{d t}= - \lambda m_{\theta_h z_k}+ m_{z_k v_h} - \mu
\, m_{\theta_h z_k} + m_{\theta_h v_k}
\label{eq8_27}
\end{equation}
implying at equilibrium
\begin{equation}
m_{\theta_h z_k}^*=0
\label{eq_28}
\end{equation}
indicating that the dissipative degrees of fredom are uncorrelated from
the inertial ones. As regards $m_{\theta_h v_k}$ we have
\begin{equation}
\frac{ d m_{\theta_h v_k}}{ dt} = - \sum_{j=1}^3 a_{kj} \, m_{\theta_h 
\theta_j} - \sum_{j=1}^3 \gamma_{kj} \, m_{\theta_h v_j} +
\mu \,  \sum_{j=1}^3 \gamma_{kj} \, m_{\theta_h z_j} - \lambda \, m_{\theta_h v_k} + m_{v_h v_k}
\label{eq8_29}
\end{equation}
that at equilibrium reduces to
\begin{equation}
-  \sum_{j=1}^3 a_{kj} \, m_{\theta_h \theta_j}^* + m_{v_h v_k}^* =0
\label{eq8_30}
\end{equation}
Enforcing the relations previously derived and the conditions eq. (\ref{eq8_20}), the latter equation becomes in matrix form
\begin{equation}
{\bf a} \, {\bf D} = \lambda \, {\bf I}
\label{eq8_31}
\end{equation}
where ${\bf I}$ is the identity matrix. Thus,
\begin{equation}
{\bf D}= \lambda \, {\bf a}^{-1} \quad \Rightarrow  \quad 
{\bf b}= \sqrt{\lambda} \, {\bf a}^{-1/2}
\label{eq8_32}
\end{equation}
Finally,
\begin{equation}
\frac{d m_{v_h z_k}}{ dt} = - \sum_{j=1}^3 a_{hj} \, m_{\theta_j z_k}
- \sum_{j=1}^3 \gamma_{hj} m_{v_j z_k} +
\mu \, \sum_{j=1}^3 \gamma_{hj} \, m_{z_j z_k} - \mu \, m_{v_h z_k} + m_{v_h v_k} + 2 \, D_{hk}^{(c)}
\label{eq8_33}
\end{equation}
At steady-state it becomes
\begin{equation}
\mu \, \sum_{j=1}^3 \gamma_{hj} \, m_{z_j z_k}^* + m_{v_h v_k}^* + 2 \, D_{hk}^{(c)} = 0
\label{eq8_34}
\end{equation}
that, enforcing the relations previously obtained for $m_{z_j z_k}^*$
and the conditions eq. (\ref{eq8_20}), attains the form
\begin{equation}
\boldsymbol{\gamma} \, {\bf D}^{(2)} + {\bf I} + 2 \, {\bf D}^{(c)} =0
\label{eq8_35}
\end{equation}
Expressing ${\bf D}^{(2)}$ and ${\bf D}^{(c)}$ in terms of the matrices
${\bf d}$ and ${\bf c}$, eq. (\ref{eq8_35}) becomes
$\boldsymbol{\gamma} \, {\bf c}^2 + {\bf I} + {\bf d} {\bf c} + {\bf c} {\bf d}
=0$, and since ${\bf d}=\boldsymbol{\gamma}^{1/2}$, ${\bf c}$ is  the solution
of the matrix equation
\begin{equation}
\boldsymbol{\gamma} \, {\bf c}^2 + {\bf I} + \boldsymbol{\gamma}^{1/2} {\bf c} + {\bf c} \boldsymbol{\gamma}^{1/2} =0
\label{eq8_36}
\end{equation}
and thus
\begin{equation}
{\bf c}= - \boldsymbol{\gamma}^{-1/2}
\label{eq8_37}
\end{equation}
that completes the determination of the representation of the
stochastic contribution.
This result can be generalized to an arbitrary number of modes, i.e.,
to
\begin{equation}
\frac{d {\bf v}(t)}{d t} = - \sum_{h=1}^{N_i} {\bf a}^{(h)} \, e^{-\lambda_h t}
* {\bf v}(t) - \sum_{\alpha=1}^{N_i} \boldsymbol{\gamma}^{(\alpha)} \, {\bf v}(t) + \sum_{\alpha=1}^{N_i} \mu_\alpha \, \boldsymbol{\gamma}^{(\alpha)} \, e^{-\mu_\alpha \, t} * {\bf v}(t) + {\bf R}(t)
\label{eq8_38}
\end{equation}
The corresponding representation in terms of dissipative/inertial degrees 
of freedom takes the form
\begin{eqnarray}
\frac{d {\bf v}}{d t} & =  &- \sum_{h=1}^{N_d} {\bf a}^{(h)} \, \boldsymbol{\theta}^{(h)} - \sum_{\alpha=1}^{N_i} \boldsymbol{\gamma}^{(\alpha)}  \, {\bf v}
+ \sum_{\alpha=1}^{N_i}
\mu_\alpha \, \boldsymbol{\gamma}^{(\alpha)} \, {\bf z}^{(\alpha)} +
\sum_{\alpha=1}^{N_i} {\bf d}^{(\alpha)} \, \boldsymbol{\xi}^{\prime, (\alpha)}
\nonumber \\
\frac{d \boldsymbol{\theta}^{(h)}}{d t} &= &- \lambda_h \, \boldsymbol{\theta}^{(h)}
+ {\bf v} +  {\bf b}^{(h)} \, \boldsymbol{\xi}^{(h)}
\label{eq8_39} \\
\frac{ d {\bf z}^{(\alpha)}}{d t}  &= &- \mu_\alpha \, {\bf z}^{(\alpha)} + {\bf v} + 
{\bf c}^{(\alpha)} \, \boldsymbol{\xi}^{\prime, (\alpha)} \nonumber
\end{eqnarray}
with $\langle \xi^{(h)}_i(t) \xi^{(k)}_j(t^\prime) \rangle =  \delta^{hk} \,
\delta_{ij} \, \delta(t-t^\prime)$, $\langle \xi^{\prime,(\alpha)}_i(t) \xi^{\prime, (\beta)}_j(t^\prime) \rangle =  \delta^{\alpha\beta} \,
\delta_{ij} \, \delta(t-t^\prime)$, $\langle \xi^{(h)}_i(t)
 \xi^{\prime,(\alpha)}_j(t^\prime) \rangle =0$. The fluctuational
intensity matrices in this general case are given by
\begin{equation}
{\bf b}^{(h)}= \sqrt{\lambda_h} \, \left ( {\bf a}^{(h)} \right )^{-1/2}
\, , \quad {\bf d}^{(\alpha)} = \left ( \boldsymbol{\gamma}^{(\alpha)} \right )^{1/2} \, , \quad {\bf c}^{(\alpha)} = - \left ( \boldsymbol{\gamma}^{(\alpha)}
\right )^{-1/2}
\label{eq8_40}
\end{equation}
generalizing the previous expression
eq. (\ref{eq7_16})
obtained in the scalar case.

\section{Examples}
\label{app2}
 
\begin{figure}
\includegraphics[width=10cm]{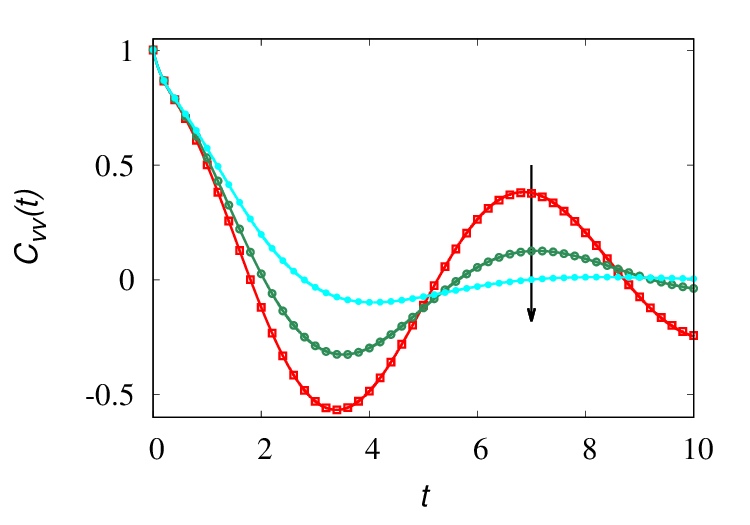}
\caption{Velocity autocorrelation function $C_{vv}(t)$ vs $t$ for model A. Symbols
represent the results of stochastic simulations, solid lines correspond to the solution of
the LRT equations (\ref{eq10_1}). The
arrow indicates increasing values of the  dissipative relaxation rate $\lambda_1=0.2,\, 0.5, \, 1$.}
\label{Fig3fd}
\end{figure}

In this paragraph, some illustrative numerical examples of the theory are presented by considering
the general case of memory kernels expressed by   eq. (\ref{eq3_1}). 
The stochastic differential equations (\ref{eq7_3}), with the values of the  expansion
coefficients given by eq. (\ref{eq7_16}),
 have been solved numerically using an Euler-Langevin 
algorithm with a time step $\Delta t=10^{-4}$.
The correlation functions have been evaluated 
considering an ensemble of $10^6$ realizations.
The stochastic results are compared with the  correlation functions deriving from
LRT,  i.e.
by solving the system of ordinary diferential equations
\begin{eqnarray}
\frac{d C_{vv}(t)}{d t} &= & - \left ( \sum_{\alpha=1}^{N_i} \right ) \, C_{vv}(t)
- \sum_{h=1}^{N_d} a_h C_{\theta_h v}(t) + \sum_{\alpha=1}^{N_i} \gamma_\alpha \, \mu_\alpha \, C_{z_\alpha,v}(t) \nonumber \\
\frac{d C_{\theta_h v}(t)}{d t} & = & - \lambda_h \, C_{\theta_h v}(t)+ C_{vv}(t)
\label{eq10_1} \\
\frac{d C_{z_\alpha v}(t)}{ d t}&= & - \mu_\alpha \, C_{z_\alpha v}(t) + C_{vv}(t)
\nonumber
\end{eqnarray}
starting from  the initial condition $C_{vv}(0)=1$, $C_{\theta_h v}(0)=0$, $h=1,\dots,N_d$,
$C_{z_\alpha v}(0)=0$, $\alpha=1,\dots,N_i$. By solving eqs. (\ref{eq10_1}),
 the behaviour
of the correlation functions $C_{\theta_h v}(t)$, $C_{z_\alpha v}(t)$ is determined in passing.
To begin with, consider the  case (model A) with
 $N_d=N_i=1$, $a_1=1$, $\gamma_1=1$, and $\mu_1=5$ keeping
$\lambda_1$ as a parameter. The velocity autocorrelation function  is
depicted in figure \ref{Fig3fd}. As expected, for low values of $\lambda_1$, the velocity autocorrelation
function displays significant oscillations, that are progressively  attenuated as $\lambda_1$
increases.
In model  B,  $N_d=N_i=1$, $a_1=1$, $\lambda_1=1$, $\gamma_1=1$, and the influence  of $\mu_1$
is considered. Sensible deviations from  a monotonic relaxation 
are observed for higher values of $\mu_1$. 

\begin{figure}
\includegraphics[width=10cm]{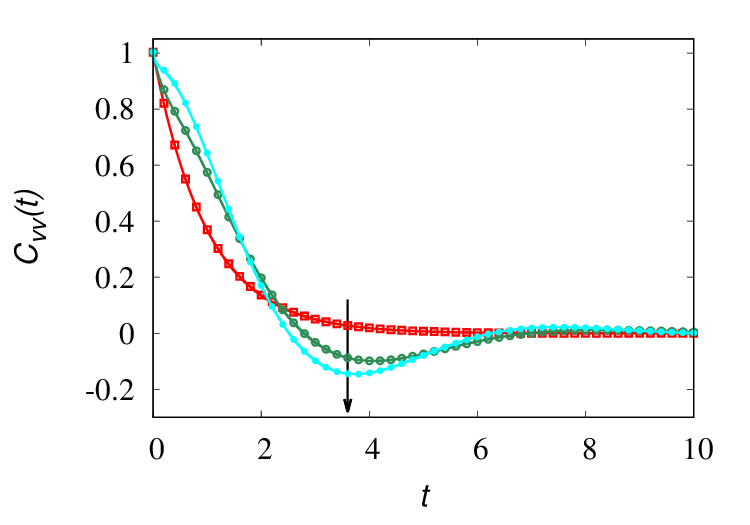}
\caption{Velocity autocorrelation function $C_{vv}(t)$ vs $t$ for model B. Symbols
represent the results of stochastic simulations, solid lines correspond to the solution of
the LRT equations (\ref{eq10_1}). The
arrow indicates increasing values of the  inertial relaxation rate $\mu_1=1,\, 5, \, 2$.
}
\label{Fig4fd}
\end{figure}
\begin{figure}
\includegraphics[width=9cm]{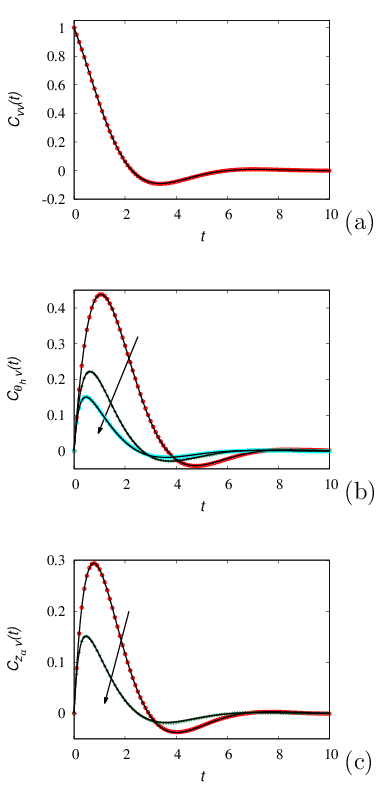}
\caption{Spectrum of correlation functions $C_{vv}(t)$, $C_{\theta_hv}(t)$ and $C_{z_\alpha v}(t)$
 vs $t$ for model C. Symbols
represent the results of stochastic simulations, solid lines correspond to the solution of
the LRT equations (\ref{eq10_1}). 
Panel (a) refers to $C_{vv}(t)$. Panel (b) to $C_{\theta_h v}(t)$, and the arrow
indicates increasing values of $h=1,\dots,N_d$. Panel (c) to $C_{z_\alpha v}(t)$,
and the arrow indicates increasing values of $\alpha=1,\dots,N_i$.}
\label{Fig5fd}
\end{figure}

In both cases, a perfect agreement between stochastic simulations and LRT can be observed,
consistently with the correctness of the modal representation for the  FD2k-FD3k
relations.
Finally, figure \ref{Fig5fd} depicts the whole spectrum of correlation
functions in model C, in which $N_d=3$, $N_i=2$, $a_1=1$, $a_2=0.5$, $a_3=0.3$,
$\lambda_1=1$, $\lambda_2=3$, $\lambda_3=5$, $\gamma_1=0.3$, $\gamma_2=0.2$, $\mu_1=2$, $\mu_2=5$.  Observe that in this case the normalization
condition eq. (\ref{eq3_3})  is not satisfied, but this is immaterial
in the present analysis.

Also in this case the agreement between theory and simulations is excellent.
The latter case represents a toy model of the typical situation occurring in complex fluids.
In general, the viscoelastic response of a complex fluid can be accurately described
by a relatively small number $N_d \simeq 1 \div 10$ of modes \cite{makosko}.
Slightly different is the analysis of fluid inertial effects: owing to presence
of the Basset $1/\sqrt{t}$-effect induced by the Laplacian  operator entering the
linearized Stokes equations,  albeit regularized by the effects of viscoelasticity \cite{gionavisco},
the number $N_i$ of fluid inertial (exponential) modes required for an accurate representation
of the kernel $k(t)$ is $N_i= k \, N_d$, with $k \simeq 10 \div 100$,
i.e. significantly larger than $N_d$. This does not represent an issue in the present theory,
but it makes solely the stochastic simulations more time consuming, as not only the
number of inertial modes increases, but  also the time-step in the simulations should be
carefully chosen in order to account for a potentially large spectrum of relaxation
rates.

\section{Fluctuation-dissipation in confined geometries}
\label{app3}

First of all, consider eqs. (\ref{eqco14}). Eq. (\ref{eqco15}) is
an elementary corollary of the principle of local consistency.
Integrating the second system of equations (\ref{eqco14}), neglecting
as irrelevant the stochastic perturbations, and assuming the particle
position ${\bf x}$ as a constant parameter, we
have
\begin{equation}
{\bf z}_\alpha(t)= e^{-\mu_\alpha \, t} {\bf z}_\alpha(0)
+ \boldsymbol{\zeta}_\alpha({\bf x}) \, e^{-\mu_\alpha \, t} * {\bf v}(t) \, ,
\alpha=1,\dots,N_i
\label{eqapp31}
\end{equation}
The exponentially decaying initial contribution can be neglected and the
substitution of eq. (\ref{eqapp31}) within eq.  (\ref{eqco14})  yields
\begin{equation}
m \frac{d {\bf v}(t)}{d t}= - \boldsymbol{\eta}({\bf x}) \, {\bf v}(t)
- {\bf G}({\bf x}) \, {\bf v}(t) + \sum_{\alpha=1}^{N_i} \mu_\alpha
\, \boldsymbol{\gamma}_\alpha({\bf x}) \, \boldsymbol{\zeta}_\alpha({\bf x})
\, e^{-\mu_\alpha \, t} * {\bf v}(t)
\label{eqapp32}
\end{equation}
from which,  by comparison with eq. (\ref{eqco13}),
the constraints eq. (\ref{eqco15}) follow.

Next, analyze the dynamics eq. (\ref{eqco15}), still considering the particle
position ${\bf x}$ as a parameter.
Componentwise, it reads as
\begin{eqnarray}
\frac{d v_h(t)}{d t} &=  & - \sum_{k=1}^3 \frac{\eta_{hk}({\bf x}) \, v_k(t)}{m} - \sum_{k=1}^3 \frac{G_{hk}({\bf x}) \, v_k(t)}{m} 
+ \sum_{\alpha=1}^{N_i} \sum_{k=1}^3 \frac{\gamma_{\alpha,hk}({\bf x}) \, \mu_\alpha \, z_{\alpha,k}(t)}{m}  \nonumber \\
&+ & \sqrt{2}  \sum_{k=1}^3 \frac{a_{hk}({\bf x}) \, \xi_k(t)}{m}
+ \sqrt{2} \sum_{\alpha=1}^{N_i} \sum_{k=1}^3 d_{\alpha,hk}({\bf x}) \,
\xi_{\alpha,k}^\prime(t) \label{eqapp33}
\noindent \\
\frac{d z_{\alpha,h}(t)}{d t} & = & - \mu_\alpha \, z_{\alpha,h}(t)
+ \sum_{k=1}^3 \zeta_{\alpha,hk}({\bf x}) \, v_k(t) + \sqrt{2} \sum_{k=1}^3
c_{\alpha,hk}({\bf x}) \, \xi_{\alpha,k}^\prime(t)
\nonumber
\end{eqnarray}
$h=1,2,3$, $\alpha=1,\dots,N_i$, where $\eta_{hk}({\bf x})$, $\gamma_{\alpha,hk}({\bf x})$ are the
entries of the matrices $\boldsymbol{\eta}({\bf x})$ and $\boldsymbol{\gamma}_\alpha({\bf x})$, and similarly for the other matrices.

The associated Fokker-Planck equation for the probability densiy
$p({\bf v},\{{\bf z}_\alpha \}_{\alpha=1}^{N_i},t;{\bf x})$,
treating ${\bf x}$ as a parameter, is given by
\begin{eqnarray}
\frac{\partial p}{\partial t} & = &
\sum_{h=1}^3 \frac{\partial }{\partial v_h} \left ( \sum_{k=1}^3 
\frac{\eta_{hk}({\bf x}) \, v_k}{m} \, p \right ) 
+  \sum_{h=1}^3 \frac{\partial }{\partial v_h} \left ( \sum_{k=1}^3 
\frac{G_{hk}({\bf x}) \, v_k}{m} \, p \right ) 
- \sum_{h=1}^3 
\sum_{\alpha=1}^{N_i} \sum_{k=1}^3  \frac{ \gamma_{\alpha,hk}({\bf x}) \, \mu_\alpha \, z_{\alpha,k}}{m} \, \frac{\partial p}{\partial v_h}  \nonumber \\
& + &  \sum_{\alpha=1}^{N_i} 
\sum_{h=1}^3 \frac{\partial }{\partial z_{\alpha,h}} \left [
\mu_\alpha \, z_{\alpha,h} \, p \right ]
- \sum_{\alpha=1}^{N_i} \sum_{h=1}^3 \sum_{k=1}^3  
\zeta_{\alpha,hk}({\bf x}) \, v_k \,  \frac{\partial p}{\partial z_{\alpha,h}}+
\sum_{h=1}^3 \sum_{k=1}^3 
 \sigma_{hk}^o({\bf x}) \, \frac{\partial^2 p}{\partial v_h \partial v_k }
\nonumber \\
&+&  \sum_{\alpha=1}^{N_i}  \sum_{h=1}^3 \sum_{k=1}^3 \sigma_{\alpha,hk}({\bf x}) 
\,  \frac{\partial^2 p}{\partial z_{\alpha,h} \partial z_{\alpha,k}}
+ \sum_{\alpha=1}^{N_i} \sum_{h=1}^3 \sum_{k=1}^3 \frac{(d_{\alpha}^2)_{hk}({\bf x})}{m^2} \frac{\partial^2 p}{\partial v_h \partial v_k}
\nonumber \\
&+ & \sum_{\alpha=1}^{N_i} \sum_{h=1}^3 \sum_{k=1}^3 
\left ( \frac{\omega_{\alpha,hk}({\bf x})}{m} 
\right ) \frac{\partial^2 p}{\partial v_h \partial
z_{\alpha,k}}
 =  {\mathcal L}_{\bf x}[p]
\label{eqapp34}
\end{eqnarray}
where $(d_{\alpha}^2)_{hk}$ are the entries of the second power
${\bf d}_\alpha^2$ of the matrix ${\bf d}_\alpha$, and
\begin{equation}
\sigma_{hk}^o({\bf x}) = \sum_{j=1}^3 \frac{a_{hj}({\bf x}) \, a_{kj}({\bf x})}{m^2} \, , \quad
\sigma_{\alpha,hk}= \sum_{j=1}^3 c_{\alpha,hj}({\bf x}) \,
c_{\alpha,kj}({\bf x})
\label{eqapp35}
\end{equation}
\begin{equation}
\omega_{\alpha,hk}({\bf x}) = \sum_{j=1}^3 
\left [ d_{\alpha,hj}({\bf x}) c_{\alpha,kj}({\bf x})+ c_{\alpha,hj}({\bf x}) 
d_{\alpha,kj}({\bf x}) \right ] 
\label{eqapp36a}
\end{equation}
Consider the dynamics of the second-order moments,
\begin{eqnarray}
\frac{d m_{v_p v_q}}{d t} & = & - \sum_{k=1}^3 	\frac{\eta_{pk}({\bf x})}{m} m_{v_k v_q}- \sum_{k=1}^3  \frac{\eta_{qk}({\bf x})}{m} m_{v_k v_p} -
\sum_{k=1}^3  \frac{G_{pk}({\bf x})}{m} m_{v_k v_q}- \sum_{k=1}^3  \frac{G_{qk}({\bf x})}{m} m_{v_k v_p} \nonumber \\
& + & \sum_{\alpha=1}^{N_i} \sum_{k=1}^3 \frac{\gamma_{\alpha,pk}({\bf x})  \, \mu_\alpha}{m} m_{v_q z_{\alpha,k}} +  \sum_{\alpha=1}^{N_i} \sum_{k=1}^3 \frac{\gamma_{\alpha,qk}({\bf x}) \, \mu_\alpha}
{m} m_{v_p z_{\alpha,k}}  \nonumber \\
&+ & 2 \, \sigma_{p1}^o({\bf x}) + 2 \, \sum_{\alpha=1}^{N_i} \frac{(d_\alpha^2)_{pq}}{m^2}
\label{eqapp36}
\end{eqnarray}
\begin{eqnarray}
\frac{d m_{v_p z_{\alpha,q}}}{d t} & = &  - \sum_{k=1}^3  \frac{\eta_{pk}({\bf x})}{m} m_{v_k z_{\alpha,q}} - 
\sum_{k=1}^3  \frac{G_{pk}({\bf x})}{m} m_{v_k z_{\alpha,q}} + 
\sum_{\beta=1}^{N_i} \frac{\gamma_{\beta,pk}({\bf x}) \, \mu_\alpha}{m } \, 
m_{z_{\alpha,q} z_{\beta,k}} \nonumber \\
& - & 	\mu_\alpha \, m_{v_p z_{\alpha,q}} + \sum_{k=1}^3 \zeta_{\alpha,qk}({\bf x}) \, m_{v_k v_p} + \frac{\omega_{\alpha,pq}({\bf x})}{m}
\label{eqapp37}
\end{eqnarray}
\begin{eqnarray}
\frac{d m_{z_{\alpha,p} z_{\beta,q}}}{d t} & = &  - (\mu_\alpha+ \mu_\beta)
\,  m_{z_{\alpha,p} z_{\beta,q}} + \sum_{k=1}^3 \zeta_{\alpha,pk}({\bf x}) \,
m_{v_k z_{\beta,q}} + \sum_{k=1}^3 \zeta_{\beta,qk}({\bf x}) \,
m_{v_k z_{\alpha,p}} \nonumber \\
&+ & 2 \, \sigma_{\alpha,pq}({\bf x}) \, \delta_{\alpha,\beta
}
\label{eqapp38}
\end{eqnarray}
and their values at steady-state (equilibrium).
Imposing the detailed Langevin conditions 
\begin{equation}
m_{v_h z_{\alpha,k}}=0 \, , \quad m_{v_h v_k}= \frac{k_B \, T}{m} \delta_{hk}
\, , \quad h,k=1,2,3\,, \;\; \alpha=1,\dots,N_i
\label{eqapp39}
\end{equation}
we obtain from eq. (\ref{eqapp36}) enforcing the symmetry of the
matrices,
\begin{equation}
\frac{k_B \, T}{m^2} \left [ \eta_{pq}({\bf x})+ G_{pq}({\bf x})
\right ]  = \frac{k_B \, T}{m^2} \left [  \eta_{pq}({\bf x})+
\sum_{\alpha=1}^{N_i} g_{\alpha,pq}({\bf x}) \right ]
= \sigma_{pq}^o({\bf x}) + \sum_{\alpha=1}^{N_i} 
\frac{(d_\alpha^2)_{pq}({\bf x})}{m^2}
\label{eqapp310}
\end{equation}
Separating the effects of dissipation from fluid inertia, as well
as the modal contributions in the inertial  term, we get in
matrix form
\begin{equation}
m^2 \, \boldsymbol{\sigma}^o({\bf x}) = {\bf a}({\bf x})= k_B \, T \, 
\boldsymbol{\eta}({\bf x}) \quad \Rightarrow
\quad {\bf a}({\bf x}) = \sqrt{k_B \, T} \, \boldsymbol{\eta}^{1/2}({\bf x})
\label{eqapp311}
\end{equation}
\begin{equation}
k_B \, T \, {\bf g}_\alpha({\bf x}) = {\bf d}_\alpha^2({\bf x})
\quad \Rightarrow \quad {\bf d}({\bf x}) = - \sqrt{k_B \, T} \, {\bf g}_\alpha^{1/2}({\bf x})
\label{eqapp312}
\end{equation}
where  we have  chosen the negative
determination  of the square root (as discussed in the main text this choice is
indeed immaterial).

From eq.  (\ref{eqapp38}) we get 
\begin{equation}
m_{z_{\alpha,p} z_{\beta,q}}= \frac{\sigma_{\alpha,pq}({\bf x})}{\mu_\alpha}
\, \delta_{\alpha \beta}
\label{eqapp313}
\end{equation}
Substituting these results into eq. (\ref{eqapp37}) we obtain in matrix
form, omitting for notational simplicity the dependence on the parameter 
${\bf x}$,
\begin{equation}
\boldsymbol{\gamma}_\alpha \, {\bf c}_\alpha + ({\bf d}_\alpha \, {\bf c}_\alpha+ {\bf c}_\alpha \, {\bf d}_\alpha) + k_B \, T \boldsymbol{\zeta}_\alpha =0
\label{eqapp314}
\end{equation}
i.e., 
\begin{equation}
{\bf c}_\alpha^2 - \sqrt{k_B \, T} \, \left (
\boldsymbol{\gamma}_\alpha ^{1/2}\, \boldsymbol{\zeta}_\alpha^{1/2}
 \, {\bf c}_\alpha
+ \boldsymbol{\gamma}_\alpha ^{-1} \, {\bf c}_\alpha \boldsymbol{\gamma}_\alpha ^{1/2}\, \boldsymbol{\zeta}_\alpha^{1/2} \right ) +
k_B \, T \, \boldsymbol{\gamma}_\alpha^{-1} \, \boldsymbol{\zeta}_\alpha =0
\label{eqapp315}
\end{equation}
Assuming that ${\bf c}_\alpha$ commute with $\boldsymbol{\gamma}_\alpha$
(this property will be demonstrated a posteriori), we have
\begin{equation}
{\bf c}_\alpha^2 - \sqrt{k_B \, T} \, \left (
\boldsymbol{\gamma}_\alpha ^{1/2}\, \boldsymbol{\zeta}_\alpha^{1/2} 
 \, {\bf c}_\alpha
+  {\bf c}_\alpha \boldsymbol{\gamma}_\alpha ^{-1/2}\, \boldsymbol{\zeta}_\alpha^{1/2} \right ) +
k_B \, T \, \boldsymbol{\gamma}_\alpha^{-1} \boldsymbol{\zeta}_\alpha =0
\label{eqapp316}
\end{equation}
that reduces to $({\bf c}_\alpha - \sqrt{k_B T} \, \boldsymbol{\gamma}_\alpha ^{-1/2}\, \boldsymbol{\zeta}_\alpha^{1/2} )^2 ={\bf 0}$, and thus
\begin{equation}
{\bf c}_\alpha({\bf x})= \sqrt{k_B \, T} \, \boldsymbol{\gamma}_\alpha ^{-1/2}({\bf x}) \, \boldsymbol{\zeta}_\alpha^{1/2}({\bf x})
\label{eqapp317}
\end{equation}
Eqs. (\ref{eqapp311})-(\ref{eqapp312}), (\ref{eqapp317}) solve
the fluctuation-dissipation problem,  when ${\bf x}$ is treated as a parameter.
But  there is a particular choice for which all the statistical
properties of the steady-state solution do not depend on ${\bf x}$.
From eq. (\ref{eqapp312}) and from the definition eq. (\ref{eqapp35})
of the matrices $\boldsymbol{\sigma}({\bf x})$, the second-order
moments of the auxiliary ${\bf z}_\alpha$ variables are position
independent if and only if the matrices ${\bf c}_\alpha({\bf x})$
do not depend on ${\bf x}$. Resorting to eq. (\ref{eqapp317}),
this occurs if and only if the  matrix product
$\boldsymbol{\gamma}_\alpha ^{-1/2}({\bf x}) \, \boldsymbol{\zeta}_\alpha^{1/2}({\bf x})$ is a constant matrix, that without loss of generality, can be
taken equal to the identity matrix. In this way, the commutation
property of ${\bf c}_\alpha$  used above is surely satisfied.
Thus, we arrive at the important result that, assuming 
$\boldsymbol{\gamma}_\alpha({\bf x})$ and $\boldsymbol{\zeta}_\alpha({\bf x})$
equal to each other, i.e. eq. (\ref{eqco17}), we have
\begin{equation}
{\bf c}_\alpha= \sqrt{k_B \, T} \, {\bf I} \, , 
\qquad
m_{z_{\alpha,p} z_{\beta,q}}= \frac{k_B \, T}{m} \delta_{pq} \, \delta_{\alpha \beta}
\label{eqapp318}
\end{equation}
It follows that within this choice, i.e.  enforcing eq. (\ref{eqco17}), the
steady-state density  $p^*({\bf v}, \{{\bf z}_{\alpha} \}_{\alpha=1}^{N_i})$
solution of eq. (\ref{eqapp34} does not depend on ${\bf x}$
and is given by
\begin{equation}
p^*({\bf v}, \{{\bf z}_{\alpha} \}_{\alpha=1}^{N_i})=
C \, \exp \left ( - \frac{m |{\bf v}|^2}{2 	\, k_B \, T}  \right )
\, \prod_{\alpha=1}^{N_i} \exp \left ( - \frac{ \mu_\alpha |{\bf z}_\alpha|^2}{2   \, k_B \, T}  \right )
\label{eqapp319}
\end{equation}
where $C$ is the normalization constant.

Next, consider the  real dynamic problem where ${\bf x}(t)$ evolves
as a function of time according to the kinematic equation in a closed
and bounded domain $\Omega$. 
The  probability density function $p({\bf x},{\bf v},\{{\bf z}_\alpha \}_{\alpha=1}^{N_i}, t)$ is the solution of the Fokker-Planck equation
\begin{equation}
\frac{\partial p({\bf x},{\bf v},\{{\bf z}_\alpha \}_{\alpha
=1}^{N_i}, t)}{\partial t} = - {\bf v} \cdot \nabla_{\bf x}
p({\bf x},{\bf v},\{{\bf z}_\alpha \}_{\alpha
=1}^{N_i}, t)+ {\mathcal L}_{\bf x}[p({\bf x},{\bf v},\{{\bf z}_\alpha \}_{\alpha
=1}^{N_i}, t)]
\label{eqapp320}
\end{equation}
where  the differential operator ${\mathcal L}_{\bf x}$ is defined 
by eq. (\ref{eqapp34}).
It is straightforward to see that eq. (\ref{eqapp319}) is the steady-state
solution of eq. (\ref{eqapp320}) as $p^*({\bf v}, \{{\bf z}_{\alpha} \}_{\alpha=1}^{N_i})$ does not depend on ${\bf x}$ and, by linearity, it is indeed the
unique equilibrium solution.   This proves Theorem II, while
Theorem III is just a consequence of Theorem II, i.e. of eq. (\ref{eqco16})
enforcing the conditions eq. (\ref{eqco17}).

To conclude, consider the case of a position-dependent added mass tensor
${\bf m}_a({\bf x})$, that is symmetric and positive definite
\cite{landau,std3}. Set,
\begin{equation}
m \, \boldsymbol{\omega}({\bf x})= m \, {\bf I} + {\bf m}_a({\bf x})
\label{eqaa1}
\end{equation}
The dynamic equations  of particle motion  attain still
the form eqs. (\ref{eqco11})-(\ref{eqco12}) with $\boldsymbol{\eta}({\bf x})$
and ${\bf g}_\alpha({\bf x})$, $\alpha=1,\dots,N_i$ substituted by
$\widetilde{\boldsymbol{\eta}}({\bf x})$ and 
$\widetilde{\bf g}_\alpha({\bf x})$, where
\begin{equation}
\widetilde{\boldsymbol{\eta}}({\bf x}) = \boldsymbol{\omega}^{-1}({\bf x})
\, \boldsymbol{\eta}({\bf x}) \, ,
\quad
\widetilde{\bf g}_\alpha({\bf x}) = \boldsymbol{\omega}^{-1}({\bf x})
\, {\bf g}_\alpha({\bf x}) \, , \;\; \alpha=1,\dots,N_i
\label{eqaa2}
\end{equation}
Consequently, the theory applies also to this case.

\end{document}